\documentstyle[12pt]{report}

\newcommand{\be}{\begin{equation}}
\newcommand{\e}{\end{equation}}
\newcommand{\la}{\lambda}
\newcommand{\vi}{\varphi}
\newcommand{\ww}{{\cal W}}
\newcommand{\is}{\int\!\!\!\!\!\!\!\mbox{$\sum$}}
\newcommand{\isf}{\int_f\!\!\!\!\!\!\!\mbox{$\sum$}}
\newcommand{\mif}[2]{\mbox{$\frac{#1}{#2}$}}
\newcommand{\dfrac}[2]{\frac{\D #1}{\D #2}}
\newcommand{\bm}[1]{\mbox{\boldmath$ #1 $}}
\newcommand{\ie}{\iota_\epsilon}
\newcommand{\II}{I\!\!\!\:I}
\newcommand{\D}{\displaystyle}

\newcommand{\back}{\!\!\!\!\!\!\!\!\!\!\!\!\!\!\!\!\!\!\!\!}
\newcommand{\bac}{\!\!\!\!\!\!\!\!\!\!}
\newcommand{\slq}{q\!\!\!\!\:/}
\newcommand{\fig}{fig.\hspace{.1cm}}

\begin{document}
\setlength{\baselineskip}{20pt}

{
\thispagestyle{empty}
\evensidemargin0cm
\oddsidemargin0cm
\vspace*{2.7cm}
\begin{center}
\renewcommand{\baselinestretch}{1.5}
{\LARGE\bf  The Electroweak Phase Transition}
\end{center}
\vspace*{4.0cm}
\begin{center}
{\large DISSERTATION} \\
\end{center} { \begin{center}
\baselineskip 0.8cm
           zur Erlangung des Doktorgrades \\
           des Fachbereichs Physik \\
           der Universit\"at Hamburg
\end{center}
\vspace*{1.0cm}
\begin{center}
\baselineskip 0.8cm
     vorgelegt von \\
     ARTHUR HEBECKER\\
     aus Moskau\\
\end{center}
\vspace*{1.5cm}
\begin{center}
\baselineskip 0.8cm
     Hamburg \\
     1995 \\
\end{center}}
\newpage
\thispagestyle{empty}
\vspace*{14cm}
\begin{tabular}{ll}
Gutachter der Dissertation: & Prof.~Dr.~W.~Buchm\"uller\\
                            & Prof.~Dr.~G.~Mack\\
                            & Prof.~Dr.~C.~Wetterich\\ \\
Gutachter der Disputation:  & Prof.~Dr.~W.~Buchm\"uller\\
                            & Prof.~Dr.~J.~Bartels\\ \\ \\
Datum der Disputation:      & 2. Mai 1995\\ \\ \\
Sprecher des Fachbereichs   & \\
Physik und Vorsitzender     & \\
des Promotionsausschusses:  & Prof.~Dr.~B.~Kramer\\
\end{tabular}
}
\thispagestyle{empty}
\mbox{}
\newpage

%\vspace*{3cm}
%\begin{center}
%{\LARGE\bf The Electroweak Phase Transition}
%\vspace{2cm}
%
%{\large (Draft)}
%\vspace{3cm}
%
%A. Hebecker
%\end{center}
%\thispagestyle{empty}
%\newpage

\thispagestyle{empty}
\vspace*{0cm}
\begin{center}
{\bf Abstract}
\end{center}
The electroweak phase transition is investigated by means of the perturbatively
calculated high temperature effective potential. An analytic result to order
$g^4,\la^2$ is presented for the Abelian Higgs model, the SU(2)-Higgs model and
the standard model and a complete on-shell renormalization at zero temperature
is performed. Higher order corrections are found to increase the strength of
the first order phase transition in the non-Abelian model, opposite to the
Abelian case. This effect is traced back to the infrared contributions from the
typical non-Abelian diagrams. The dependence of several phase transition
parameters on the Higgs mass is analysed in detail. A new, gauge invariant,
approach based on the composite field $\Phi^\dagger\Phi$ is introduced. This
method, which supports the above Landau gauge results numerically, permits a
conceptually simpler treatment of the thermodynamics of the phase transition.
In particular, it enables a straightforward comparison with lattice data and
the application of the Clausius-Clapeyron equation to the electroweak phase
transition.

\vspace*{1.4cm}
\begin{center}
{\bf Zusammenfassung}
\end{center}
Der elektroschwache Phasen\"ubergang wird mit Hilfe des st\"orungstheoretisch
be\-rech\-ne\-ten effektiven Potentials der Hochtemperaturfeldtheorie
untersucht. F\"ur das abelsche Higgs-Modell, das SU(2)-Higgs-Modell und das
Standardmodell werden analytische Ergebnisse der Ordnung $g^4,\la^2$ angegeben,
und eine voll\-st\"an\-di\-ge Renormierung im On-Shell-Schema bei Temperatur
Null wird durchgef\"uhrt. Im Gegensatz zum abelschen Fall f\"uhren die
Kor\-rek\-tu\-ren h\"oherer Ordnung beim nicht\-abel\-schen Modell zu einer
Verst\"arkung des Pha\-sen\-\"uber\-gan\-ges erster Ordnung. Dieser Effekt wird
auf die infraroten Beitr\"age typischer nicht\-abel\-scher Dia\-gram\-me
zur\"uckgef\"uhrt. Die Abh\"angigkeit mehrerer Parameter des
Pha\-sen\-\"uber\-gan\-ges von der Higgs-Masse wird einer detaillierten Analyse
unterzogen. Ein neuer, eichinvarianter Zugang, der sich auf das
zusammengesetzte Feld $\Phi^\dagger\Phi$ st\"utzt, wird eingef\"uhrt. Diese
Me\-tho\-de best\"atigt numerisch die obigen, in Landau-Eichung erzielten
Resultate und erlaubt eine begriff\/lich einfachere Behandlung der
Thermodynamik des Phasen\"uberganges. Insbesondere erm\"oglicht sie den
unmittelbaren Vergleich mit Gitter-Daten und die Anwendung der
Clausius-Clapeyron-Gleichung auf den elektroschwachen Phasen\"ubergang.
\newpage
\thispagestyle{empty}
\mbox{}
\newpage
\pagenumbering{arabic}
\tableofcontents

\chapter*{Introduction}
\addcontentsline{toc}{chapter}{Introduction}
The whole known universe consists almost exclusively of matter, with no
considerable amount of antimatter in our galaxy cluster and no known mechanism
to separate matter and antimatter on such large scales \cite{KT}. This baryon
asymmetry of the universe is one of the most interesting cosmological problems
to be resolved by particle theory.

As pointed out by Sakharov in 1967 the baryon asymmetry may be a calculable
result of particle interactions \cite{Sa}, the necessary conditions being
baryon number violation, C-- and CP--violation, and departure from thermal
equilibrium.

Kirzhnits and Linde realized in 1972 that at high temperatures the
spontaneously broken electroweak symmetry is restored, thus suggesting a phase
transition in the early universe \cite{KL}. Since anomalous baryon number
violation in the standard model is rapid at high temperatures, the departure
from equilibrium in a first order electroweak phase transition opens the
possibility of standard model baryogenesis. This scenario, first suggested by
Kuzmin, Rubakov and Shaposhnikov in 1985 \cite{KRS}, provides the main
motivation for the present investigation. Although more recent analyses seem to
discourage baryogenesis within the minimal standard model due to the small
CP-violation, simple non-minimal models may produce a sufficient asymmetry (see
\cite{BG} and references therein). In any case it is clear that the present
baryon asymmetry of the universe has been finally determined at the electroweak
phase transition, since baryon-number violating processes fall out of thermal
equilibrium at the corresponding critical temperature.

A quantitative
understanding of the electroweak phase transition is a basic
prerequisite for the discussion of any model of baryogenesis at the weak scale.
This includes reliable knowledge of its order and of the strength of the phase
transition, if it is of first order. Several approaches have been used to
investigate the electroweak phase transition. Important results have been
obtained by use of 3-dimensional effective theory \cite{PATKOS,FKRS},
$\epsilon$-expansion \cite{EE}, average action \cite{RW} and lattice
simulations \cite{LATTICE,L1,L2}. In particular, the detailed lattice data for
physical phase transition parameters from refs.~\cite{L1,L2} permit explicit
comparison with results from perturbation theory.

The present investigation is concerned with the extraction of thermodynamic
parameters of the electroweak phase transition, based on the perturbative
calculation of the high temperature effective potential, i.e. the free energy
of the system.

Perturbative calculations of the potential to the order $g^3,\la^{3/2}$ suggest
a first order phase transition for different Higgs models ($g$ denotes the
gauge coupling and $\la$ the scalar coupling). These calculations, based on the
one-loop ring summation, have been carried out in refs.~\cite{Ar,BHW} for the
Abelian Higgs model and in refs.~\cite{Ca,BFHW} for the standard model.
Two-loop summation has been done to order $g^4,\la$ in ref.~\cite{AE}, where
scalar masses have been neglected with respect to gauge boson masses, and by
use of another approximation in ref.~\cite{BD}. The results of ref.~\cite{AE}
include both the Abelian and the non-Abelian case.

However, there is a need to extend the work of Arnold and Espinosa \cite{AE} to
a complete $g^4,\la^2$-calculation. In the present investigation, assuming
formally $\la\sim g^2$ and keeping the full dependence on the Higgs field
$\vi$, its zero temperature vacuum expectation value $v$ and the temperature
$T$, the necessary corrections are calculated \cite{He,FH}. To obtain
information about the importance of specific non-Abelian effects and to have a
particularly simple model, the Abelian case has been considered first
\cite{He}.

Besides the above extension of a conventional approach, a gauge invariant
calculation of phase transition parameters is presented \cite{BFH,BFH1}. Using
this method a better understanding of the physics of the phase transition is
obtained. The numerical results of the gauge invariant approach are similar to
conventional Landau gauge calculations, thus supporting their reliability.
{\parskip3ex

Chapter} \ref{Veff} starts with some well known facts about the effective
potential, its thermodynamic interpretation, its loop expansion and the
relevance for the description of first order phase transitions. After that the
resummation of masses, necessary at high temperature, is discussed. Here the
emphasis is on a method based on Dyson-Schwinger equations \cite{He,FH} to be
used in the sequel.

In chapter \ref{ahm} the complete $g^4,\la^2$ calculation of the finite
temperature effective potential is performed for the Abelian Higgs model
\cite{He}. This includes a zero temperature renormalization in the on-shell
scheme. The absence of a linear $\vi$-term, explicitly verified at this order,
is shown to survive to all orders. Latent heat, surface tension and jump of the
order parameter are calculated as functions of the Higgs mass from different
approximations to the effective potential. The higher order scalar corrections,
added to the previous results, are found to be important at not too small Higgs
masses.

The above analysis is extended to the standard model in chapter \ref{sm} which
is based on ref.~\cite{FH}. First, the case of the pure SU(2)-Higgs model,
which is much simpler, is discussed in detail. The importance of specifically
non-Abelian contributions is stressed and the uncertainties of perturbation
theory are traced back to infrared problems. A qualitatively similar situation
is found for the complete standard model. The main quantitative change is
introduced by the large top mass, which reduces the strength of the first
order phase transition.

In the last chapter the gauge invariant approach to the phase transition,
suggested in ref.~\cite{BFH}, is described. The phase transition parameters
obtained from a one-loop calculation are compared with the Landau gauge
results, showing good agreement for not too large Higgs masses. Several
additional questions of the analysis of the electroweak phase transition are
discussed in the sequel \cite{BFH1}. They include two-loop resummation
problems, the connection of gauge invariant and Landau gauge approach, and the
use of the Clausius-Clapeyron equation in the present context.

After the discussion of conclusions to be drawn from the above investigation
several analytic formulae are listed in the appendices. Appendix \ref{int}
contains some useful integrals. The explicit analytic results for the finite
temperature effective potential in the Abelian Higgs model, the SU(2)-Higgs
model and the standard model are displayed in appendix \ref{results}. Self
energy corrections at zero temperature, necessary for the on-shell
renormalization, are given in appendix \ref{se}.

\chapter{The effective potential}\label{Veff}

\section{Definition and loop expansion}\label{def}
The present investigation is concerned with equilibrium thermodynamics, which,
as is well known, can be completely described as soon as the partition function
    \be Z=\mbox{\bf tr}\exp\left[-\beta(\bm{H}+\int_V J\bm{\vi})\right] \e
of the system under consideration is given. Here the inverse temperature is
denoted by $\beta$ and $\bm{H}$ is the Hamilton operator. The source $J$ is
coupled to the field $\bm{\vi}$, specified later on as the Higgs field, which
is used as the order parameter for the description of the phase transition. In
this section it is sufficient to consider the simplest possible case of a
$\vi^4$-model with one degree of freedom. The generalization to more
complicated field theories is straightforward.

The fundamental thermodynamic potential per unit volume, $W(J)$, is related to
the partition function by
    \be Z=\exp(-\beta\Omega W)\, , \e
where $\Omega$ denotes the three dimensional volume of the physical system. The
temperature dependent effective potential is now defined by the transition from
the variable $J$ to the variable $\vi$, realized by a Legendre transformation:
    \be V(\vi,T)=W(J,T)-J\vi\qquad,\qquad\vi=\frac{\partial W(J,T)}{\partial J}
    .\e
It is straightforward to derive the two identities
    \be <\bm{\vi}>=\vi=\frac{\partial W}{\partial J}\qquad,\qquad
    \frac{1}{\Omega}<\bm{H}>\Big|_{\D <\bm{\vi}>=\vi}-TS=V(\vi,T) \e
clarifying the physical interpretation of the effective potential as the free
energy of the system. Here the brackets $<...>$ symbolize the thermal
expectation value of an operator and $S=-\partial W/\partial T$ is the entropy.

The perturbative calculation developed later on is based on the path integral
representation of the partition function \cite{K}
    \be Z=\mbox{\bf tr}\exp\left[-\beta(\bm{H}+\int_V J\bm{\vi})\right]=
    \int D\vi_{\mbox{\scriptsize period.}}\exp\left[\int_0^\beta d\tau\int d^3
    \vec{x}({\cal L}-J\bm{\vi})\right]\, .\label{ZW}\e
Here ${\cal L}$ is the Euclidean Lagrangian and the path integral is taken
over all fields periodic in time direction. Compactification of Euclidean time
results in the replacement of the well known loop integral by the integral-sum
    \be \is dk=\frac{T}{(2\pi)^3}\sum_{n=-\infty}^{n=+\infty}\int d^3\vec{k}\e
and correspondingly in the introduction of the Euclidean propagator with
discrete Matsubara frequencies $k_0$
    \be \frac{1}{k^2+m^2}=\frac{1}{k_0^2+\vec{k}^2+m^2}\qquad,\qquad k_0=2\pi
    Tn.\e
This imaginary time formalism, used exclusively in the following, is the
simplest formulation of finite temperature field theory and perfectly suited
for the investigation of the desired equilibrium parameters.

While $Z$ is the sum of all Feynman diagrams, $\beta\Omega W=-\ln Z$ contains
the connected graphs only. The potential $V$, obtained from $W$ by means of a
Legendre transformation, is the momentum independent part of the generating
functional of one-particle irreducible diagrams. It has to be evaluated at
non-vanishing external field, thus requiring the summation of infinitely many
graphs at each loop order. This problem is solved by the following identity
\cite{Ja}:
    \be V(\hat{\vi})=V_{tree}(\hat{\vi})+\frac{1}{2}\is dk\ln(k^2+m^2_{\hat{\vi
    }})+\Big\{\mbox{one-particle irred. vacuum diagrams}\Big\}.\label{vacuum}\e
Here $V_{tree}$ denotes the tree-level potential, i.e. the momentum independent
part of the free Hamiltonian. The curly bracket represents the sum of all
one-particle irreducible vacuum graphs calculated from the `shifted' Lagrangian
    \be {\cal L}^{\hat{\vi}}(\vi)={\cal L}(\hat{\vi}+\vi)-\Big\{\mbox{terms,
    constant and linear in $\vi$}\Big\}\, . \label{shift}\e
Correspondingly, $m^2_{\hat{\vi}}$ is the mass term generated after this shift.

The perturbative methods described in section \ref{res} are based on the above
representation of the potential. Note that in the following sections, the shift
$\hat{\vi}$ will often be denoted by $\vi$ for brevity, if no confusion is
possible.

In the remainder of this section a short derivation of the identity
(\ref{vacuum}) shall be given. To keep the notation compact, let
$\beta=\Omega=1$ during these manipulations.

Shifting the integration variable in eq.~(\ref{ZW}) by some arbitrary but fixed
function $\tilde{\vi}$ results in
    \be W[J]=-\ln\int D\vi\exp\Bigg[\int{\cal L}(\vi+\tilde{\vi})-(\vi+
    \tilde{\vi})J\Bigg]\, ,\e
where the Lagrangian after the shift can be written in the form
    \be {\cal L}(\vi+\tilde{\vi})=-V_{tree}(\tilde{\vi})-\vi g(\tilde{\vi})+
    {\cal L}^{\tilde{\vi}}(\vi)\, .\e
Note that $J$ is not necessarily constant and $W$ is a functional. The
effective action is defined by
    \be \Gamma[\hat{\vi}]=-W[J]+\int J\hat{\vi}\qquad,\qquad
    \hat{\vi}=\frac{\delta W}{\delta J}\, ,\e
where arguments have been dropped for brevity in the functional derivative.
Introducing the notation $\tilde{J}=J+g(\tilde{\vi})$ and using the generating
functional of the `shifted' theory
    \be W^{\tilde{\vi}}[J]=-\ln\int D\vi\exp\Bigg[\int{\cal L}^{\tilde{\vi}}(
    \vi)-\vi J\Bigg]\, ,\e
the effective action takes the form
    \be \Gamma[\hat{\vi}]=-V_{tree}(\tilde{\vi})-W^{\tilde{\vi}}[\tilde{J}]+
    \int(\hat{\vi}-\tilde{\vi})J\, .\e
Observe that
    \be \hat{\vi}=\frac{\delta W[J]}{\delta J}=\frac{\delta W[J]}
    {\delta\tilde{J}}=\tilde{\vi}+\frac{\delta W^{\tilde{\vi}}[\tilde{J}]}
    {\delta\tilde{J}}\, .\e
Now the shift $\tilde{\vi}$, which has not yet been specified, is set to
$\tilde{\vi}=\hat{\vi}$. This gives an explicit result for the effective
action:
    \be \Gamma[\hat{\vi}]=-V_{tree}(\hat{\vi})-W^{\hat{\vi}}[\tilde{J}]
    \quad\mbox{at}\quad\frac{\delta W^{\hat{\vi}}}{\delta \tilde{J}}=0\, .\e
Since the effective potential is given by minus the momentum independent part
of $\Gamma$, the identity (\ref{vacuum}) follows.

\section{Description of a first order phase transition}
\begin{figure}[t]
\refstepcounter{figure}
\label{dw}\hspace{2cm}\parbox[b]{9cm}{
{\bf Fig.\ref{dw}} Schematic graph of the coarse-grained free energy, $f(c)$,
                  and the corresponding true free energy, $\tilde{f}(c)$. The
                  dotted sections denote the analytic continuation of
                  $\tilde{f}(c)$ into the metastable region
                  (from Langer, \cite{Lan})\\[-.05cm]\mbox{}
}\end{figure}

The intuitive picture of a first order phase transition, described by the free
energy density of the system, is based on the double well structure of the
free energy as a function of the order parameter. In this picture the barrier
between the two minima is responsible for the necessity of an activation energy
for a transition between the phases, thus rendering the phase transition first
order. However, a rigorous definition of such a non-convex free energy is not
trivial. First the phenomenological discussion of Langer \cite{Lan} shall be
briefly described:

Assume the existence of a coarse-grained free energy, $F(c)$, of the form
    \be F(c)=\int_V\left[\frac{1}{2}(\partial c)^2+f(c)\right]\, ,\e
with some function $f(c)$ characterizing the homogeneous state and an order
parameter $c$ of the system. The free energy density $f$, typically of the form
given in \fig\ref{dw}, is well suited for the description of the metastable and
unstable region. However, even in the region where $f$ has a positive second
derivative it is not identical with the true free energy density $\tilde{f}$.
The later one is a convex function and connects the two different physical
states of the system by a straight line. In the metastable, though not in the
unstable region, the analytic continuation of $\tilde{f}(c)$ does still
describe the thermodynamic properties of the one-phase physical system.

The perturbatively calculated high
temperature effective potential is by definition an analytic function of the
order parameter $\vi$. Therefore, assuming its convergence to the true free
energy in the stable region \cite{Sl,Wie}, it can be naturally interpreted in
the above sense as the free energy of the metastable states. This, however,
does not clarify the interpretation in the unstable, non-convex region of the
potential.

Following Langer, the coarse-grained free energy can be calculated by
integrating out the short wavelength components of the microscopic variable
only. This corresponds to the introduction of some infrared cutoff
characterizing the coarse-graining size. Note that the one-loop results for
the effective potential are not very sensitive to a small enough infrared
cutoff. In fact, in the Abelian model no infrared problems are expected at any
loop-order (see section \ref{lt}). Even the non-Abelian two-loop results of
section \ref{su2-num} do not change qualitatively when a small infrared cutoff
is introduced. This could be taken as a justification to interpret the
non-convex region of the obtained effective potential in the spirit of the
coarse-grained free energy. A detailed discussion of the coarse-grained free
energy in high temperature field theory can be found in ref.~\cite{BBFH}.

At one-loop level such an interpretation is supported by the analysis of
ref.~\cite{WW}. There it is shown that in some approximation the effective
potential $V(\vi)$ gives the energy density of a homogeneous state with wave
functional concentrated on configurations near the classical value $\vi$.

It has to be admitted that the understanding of the perturbatively calculated
effective potential in the non-convex region does not seem to be
satisfactory. This implies some doubts about the physical interpretation of
quantities calculated from the potential in that region. In particular, the
surface tension, as it is defined and calculated in chapters \ref{ahm} and
\ref{sm} is affected by this uncertainty.

Nevertheless, critical temperature, latent heat and jump of the order parameter
are, in principle, calculable by standard methods from the effective potential.
This is due to the fact that they can be obtained from the minima of the free
energy, which, following the above discussion, are trustworthily described
by the potential.

\section{Resummation}\label{res}
The loop expansion of the potential given by eq.~(\ref{vacuum}) formally
corresponds to an expansion in coupling constants. Unfortunately, due to the
effectively three dimensional integrals arising from contributions with
vanishing Matsubara frequencies in the high temperature theory terms
proportional to $(T/m)^n$ do appear. Owing to the Higgs mechanism masses are
proportional to coupling constants in the relevant theories. Therefore a formal
loop expansion does not generate the desired expansion in coupling constants.
This problem is well known and can be solved by resummation. In the following
this shall be described in detail for some unspecified theory with a generic
coupling constant $g$, to be used as the expansion parameter:

Consider a general Lagrangian with interaction terms generating 3- and
4-vertices proportional to $g^2$ and 3-vertices proportional to $g\,k_\mu$.
Here $k_\mu$ is a momentum variable, as it appears at the vertex in gauge
theories. All masses are considered to be of order $g$. Note that this
structure is suggested by the standard model Lagrangian, where the square root
of the scalar coupling $\sqrt\la$, the Yukawa coupling $g_Y$, the electroweak
gauge couplings $g_1,g_2$ and the strong gauge coupling $g_s$ play the role of
the generic coupling $g$. The Abelian Higgs model and the SU(2)-Higgs model
fit this general structure as well\cite{BHW,BFHW,H}.

Resummation now means that the leading self-energy corrections of order
$g^2T^2$ have to be added to the mass squares, thereby collecting an infinite
series of graphs with increasing powers of $(gT/m)$. Having realized that in a
systematic way, the remaining higher loop corrections are always connected with
higher orders in the couplings, which ensures a systematic expansion of the
potential \cite{BFHW}.

Working in that spirit, the main task is now to identify all graphs
contributing to the order $g^4$.

\subsection{Method based on Dyson-Schwinger equations}\label{ds}
In this subsection a method for the calculation of the effective potential
based on Dyson-Schwinger equations is presented \cite{He,FH}. A similar way of
summing the different contributions to $V$ for the $\vi^4$-theory has been
considered in ref.~\cite{BBH}.

To circumvent the combinatoric problems of resummation it is useful not to
calculate the potential $V(\vi)$ itself but its derivative with respect to the
field $\vi$, i.e. the sum of all one-particle irreducible one-point functions.
This `tadpole' method has been suggested in refs.~\cite{WKL}.

The Dyson-Schwinger equation \cite{Riv} for the effective potential has the
form
    \be-\frac{\partial}{\partial\vi}(V-V_{tree})=A+B=
    \label{ds1}\, ,\e\\[0.3cm]
where the internal lines represent all particles of the relevant theory and the
external lines stand for the shifted scalar field. The two different sorts of
blobs are full propagator and full 3-vertex respectively. Therefore the first
term has the explicit form
   \be A=\mbox{tr}\,\ww(\vi)\is\frac{dk}{k^2+m_{tree}^2+\Pi(k)}\, .\label{A1}\e
In general, mass, self-energy and vertex $\ww$ are matrices and ``tr'' denotes
the sum over the suppressed indices. The $\vi$-dependence of the mass
$m_{tree}$, introduced by the shift (\ref{shift}), is obvious.

No corrections are needed for the vertex function in the second term of
eq.~(\ref{ds1}). This can be verified by the following argument: Any correction
to that vertex corresponds to a proper three-loop contribution to the
potential. Here by {\it proper}, the absence of self-energy insertions, which
would require resummation, is meant. The three-dimensional contribution to the
three-loop graph, arising when all Matsubara frequencies vanish, results in a
$g^5$-correction. This can be seen most easily by scaling all loop momenta
according to $\vec{k}\to\vec{k}g$. Similar arguments do also prove that the
parts of three-loop diagrams with non-vanishing Matsubara frequencies do not
contribute $\vi$-dependent terms to the potential up to order $g^4$. This way
of argumentation is explained in more detail in ref.~\cite{H}. Appendix A of
ref.~\cite{BFHW} contains a proof of the sufficiency of self-energy
resummation for the order $g^3$-potential, which can be generalized to higher
orders.

The self-energy insertions needed in both terms of eq.~(\ref{ds1}) can again be
obtained from a Dyson-Schwinger equation, which, to the required leading order,
reads
    \be-\Pi(k)=-(\Pi_a(k)+\Pi_b(k))=
\label{ds2}\quad.\e\\[0.2cm]
It is easily verified that the omitted parts would result in proper three-loop
terms in the potential which do not contribute to the order $g^4$. In the
following the indices 2 and 3 denote contributions of order $g^2$ and $g^3$
respectively. The tadpole part of the self-energy can be written as
    \be \Pi_a(k)=\Pi_{a2}+\Pi_{a2}(k)+\Pi_{a3}+\cdots\quad,\quad\mbox{with}
    \quad\Pi_{a2}(0)=0\, .\label{pia}\e
Here the momentum dependent part $\Pi_{a2}(k)$ does only arise in the case of a
non-Abelian gauge theory. It is introduced by the corresponding projection
operator when calculating the longitudinal self-energy of the gauge boson
(see section \ref{na}).

The third order term in $g$ does not contribute in the case of nonzero $k_0$.
If $k_0=0$, by the above scaling argument $\vec{k}\to\vec{k}g$ only its
momentum independent part $\Pi_{a3}$ is needed.

The leading order momentum independent $g^2$-part of $\Pi_b(k)$ will be called
$\Pi_{b2}$. All these contributions to the self-energy $\Pi(k)$, needed below,
are obtained by iterating eq.~(\ref{ds2}) twice.

Using these definitions and introducing the corrected mass term $m^2$,
    \be m^2=m_{tree}^2+\Pi_{a2}+\Pi_{b2}\quad, \e
equation (\ref{A1}) can be written as
    \begin{eqnarray}\!\!\!\!\!A&\!\!\!=&\!\!\!\mbox{tr}\,\ww(\vi)\is\frac{dk}
    {k^2+m^2+\Pi_{a2}(k)+\Pi_{a3}+\Pi_b(k)-\Pi_{b2}}\nonumber\\
\!\!\!\!\!&\!\!\!=&\!\!\!\mbox{tr}\,\ww(\vi)\is dk\Bigg(\frac{1}{k^2+m^2+
    \Pi_{a2}(k)}-\frac{1}{k^2+m^2}\Pi_{a3}\frac{1}{k^2+m^2}\nonumber\\
\!\!\!\!\!&\!\!\!&\!\!\!  +\frac{1}{k^2+m^2}\Pi_{b2}\frac{1}{k^2+m^2}
    -\frac{1}{k^2+m^2}\Pi_b(k)\frac{1}{k^2+m^2}\Bigg)\,
    .\label{A2}\end{eqnarray}
Here the second equality is obtained by expanding the integrand in $g$.
Separate treatment of the parts with $k_0=0$ and $k_0\neq 0$ together with the
above scaling argument show that all contributions of order $g^4$ are taken
into account.

Inspection of the last term of eq.~(\ref{A2}) and term $B$ of eq.~(\ref{ds1})
shows that their sum is equal to the derivative $-\partial V_\ominus/\partial
\vi$, where $-V_\ominus$ represents the sum of all two-loop diagrams of the
type shown in \fig\ref{s2l}.a (setting sun diagrams). To see this, notice that
    \be \ww(\vi)=-\frac{1}{2}\frac{\partial}{\partial\vi}m^2\, ,\e
a direct consequence of the shift generating both mass terms and 3-vertices
from the interaction Lagrangian. This observation simplifies the actual
calculation significantly.

Furthermore, the contribution
    \be V_z=\int^\vi d\vi'\mbox{tr}\,\ww(\vi')\is dk
    \frac{1}{k^2+m^2}\Pi_{a3}\frac{1}{k^2+m^2}\label{Vz}\e
is treated separately. It is equal to the sum of all terms bilinear in masses
coming from two-loop diagrams of the type shown in \fig\ref{s2l}.b, i.e. the
sum of their three-dimensional parts. Note that here and in the setting sun
diagrams discussed above, masses resummed to leading order have to be used.
\\[.3cm]
%\hspace*{1.5cm}
%\psfig{width=12.8cm,file=s2l.eps}
%\\[-1.4cm]
\begin{figure}
\refstepcounter{figure}
\label{s2l}
\end{figure}
{\bf Fig.\ref{s2l}} Typical two-loop diagrams\\[.6cm]
The remaining part
    \be V_R=-\int^\vi d\vi'\,\mbox{tr}\,\ww(\vi')\is dk\left(\frac{1}{k^2+m^2+
    \Pi_{a2}(k)}+\frac{1}{k^2+m^2}\Pi_{b2}\frac{1}{k^2+m^2}\right)\label{VR1}\e
can be easily rewritten as
    \be V_R=\frac{1}{2}\int dm^2\,\mbox{tr}\,\is\frac{dk}{k^2+m^2}+\frac{1}{2}
    \,\mbox{tr}\,\is\frac{dk}{k^2+m^2}(\Pi_{a2}(k)-\Pi_{b2})\, .\label{VR2}\e
Therefore the complete potential is given by
    \be V=V_{tree}\!+V_\ominus+V_z+V_R\, .\label{V}\e
Denoting by $V_3$ the sum of the tree level potential and the $g^3$-order part
of $V_R$ and calling $V_4$ the fourth order corrections of $V_R$,
    \be V_3+V_4=V_{tree}+V_R, \e
the following final formula is obtained :
    \be V=V_3+V_4+V_\ominus+V_z. \label{final}\e
This representation, used for the explicit results in the appendix, has the
advantage of a separation of the third order part from the higher order
corrections.

\subsection{Counterterm method}\label{ctm}
Another possibility of a systematic resummation at high temperature is based on
the introduction of thermal counterterms \cite{P}. The simplest form of this
method is to add and subtract a temperature dependent mass term for every field
in the Lagrangian. Let
    \be {\cal L}={\cal L}_{kin}+{\cal L}_{mass}+{\cal L}_I\quad\mbox{with}
    \quad{\cal L}_{mass}=-\sum_i \frac{1}{2}m_{0i}^2\vi_i^2\e
be the Euclidean Lagrangian depending on several fields $\vi_i$, with kinetic
term, mass term and interaction term. The leading temperature dependent mass
corrections are given by
    \be -\delta m_i^2=-\Pi_{2i}(0)=
    \quad,\e\\[0.01cm]
where only the $g^2T^2$-pieces of the diagrams are considered. The Lagrangian
is now rewritten according to
    \[ {\cal L}={\cal L}_{kin}+{\cal L'}_{mass}+{\cal L'}_I\, ,\]\\[-1.6cm]
    \be{}\e\\[-1.6cm]
    \[{\cal L'}_{mass}=-\sum_i\frac{1}{2}(m_{0i}^2+\delta m_i^2)\vi_i^2\qquad,
    \qquad{\cal L'}_I={\cal L}_I+\frac{1}{2}\sum_i\delta m_i^2\vi_i^2\, .\]
The main point is now in the cancellation occurring between the thermal
counterterms from the new interaction Lagrangian and those contributions from
higher loop diagrams which are of low order in the couplings and require
resummation. For example, the sum of the diagrams from \fig\ref{ct3l} does not
contribute to the field dependent part of the effective potential up to order
$g^4$. Here the dot symbolizes the thermal counterterm. If the thermal
counterterm is considered as increasing the formal loop order by one, all of
these diagrams are of three-loop order. Treating the proper three-loop
diagrams, i.e. those without self-energy parts, separately, it can be shown
that the sum of all three-loop diagrams does not contribute to the desired
$g^4$-potential.\\[.2cm]
%\psfig{width=10.1cm,file=ct3l.eps}
%\\[.1cm]
\begin{figure}
\refstepcounter{figure}
\label{ct3l}
\end{figure}
{\bf Fig.\ref{ct3l}} Three-loop diagrams with self-energy parts\\[.6cm]
Arguments of that kind can be generalized to higher loop orders. Therefore, to
obtain the full $g^4$-result in this approach, the diagrams shown in
\fig\ref{ct2l} are sufficient. Here the circle symbolizes the one-loop
contribution
which is proportional to the logarithm of the determinant of the propagator. It
has to be kept in mind that the notation in this chapter is merely a generic
one, so that the lines in the diagrams stand for all the different particles of
the theory.\\[.3cm]
\begin{figure}
\refstepcounter{figure}
\label{ct2l}
\end{figure}
%\psfig{width=9.3cm,file=ct2l.eps}
%\\[.1cm]
{\bf Fig.\ref{ct2l}} Typical diagrams contributing to the $g^4$-potential in
the counterterm method.\\[.6cm]
In ref.~\cite{AE} a slight modification of this method is used. There, the
resummation is applied only to the zero Matsubara frequency modes. Thereby
many constant terms and terms which would cancel each other in the final result
are omitted from the beginning. Also, using this method, it is not necessary to
keep the dependence on the space dimension in the thermal counterterms, if
dimensional regularization is used. The details of this will not be discussed
here, because the following calculations are based on the method of section
\ref{ds}. However, the results of chapters \ref{ahm} and \ref{sm} have been
checked by use of the counterterm method.

\chapter{Abelian Higgs model}\label{ahm}
The simplest gauge theory with spontaneous symmetry breaking, the Abelian Higgs
model, is believed to exhibit the main features of the electroweak phase
transition \cite{KL2,Li}. At one-loop order the effective potential shows a
first order phase transition at high temperature, driven by a term cubic in the
vector mass in complete analogy to the standard model case \cite{Ar,BHW}. The
study of the Abelian Higgs model might prove useful for the understanding of
the electroweak phase transition, because this simple model does not suffer
from the severe infrared problems of the non-Abelian theory. No magnetic mass
is expected to arise. This does not present a problem due to the absence of the
higher-loop non-Abelian contributions, which are divergent in the symmetric
phase.

In this chapter a complete calculation up to the order $e^4,\la^2$, with gauge
coupling $e$ and scalar coupling $\la$, is presented, supplying the results
of ref.~\cite{AE} with scalar corrections. An analysis of the phase transition
parameters shows that these corrections are important if the Higgs mass is not
too small.

\section{Calculation of the potential}\label{ahm-calc}
The effective potential is calculated by expanding it in the coupling
constants, as described in section \ref{ds}. Consider the Euclidean Lagrangian
    \be {\cal L}=-\frac{1}{4}F_{\mu\nu}F_{\mu\nu}-|D_\mu\Phi|^2+\nu|\Phi|^2-\la
    |\Phi|^4\, ,\e
with
    \be D_\mu=\partial_\mu+ieA_\mu\quad\mbox{and}\quad\Phi=\frac{1}{\sqrt{2}}
    (\hat{\vi}+\vi+i\chi).\e
Using the tree level vacuum expectation value $v$ the Higgs mass term is
rewritten as $\nu=\la v^2$ and counted as order $\la$. Then the identification
with the the generic coupling $g$ of section \ref{res} reads $g\sim
e\sim\sqrt{\la}$. At this point it is necessary to describe in more detail
the resummation procedure for the vector particle \cite{BHW,K}:

The choice of the Landau gauge, which will be used throughout this
investigation, is justified by the absence of vector -- Goldstone boson mixing
together with the $\vi$-independence of the gauge. Two-loop calculations for
the three-dimensional model do also show that the convergence of the
perturbation series is best in Landau gauge \cite{La}. The bare propagator can
be given in the form
    \be D_0(k)=\frac{1}{k^2+m^2}(P_L+P_T)\, , \e
where
    \be P_{T\,\mu\nu}=\sum_{i,j=1}^3\delta_{\mu i}\left(\delta_{ij}-
    \frac{k_ik_j}{\vec{k}^2}\right)\delta_{j\nu}\quad\mbox{and}\quad
    P_{L\,\mu\nu}=\delta_{\mu\nu}-\frac{k_\mu k_\nu}{k^2}-P_{T\,\mu\nu}.\e
Here the full covariance has been broken down to SO(3), which is the symmetry
left after the specification of a rest frame intrinsic in the formalism of
thermal field theory. In ref.~\cite{BHW} the full propagator is shown to have
the form
    \be D(k)=\sum_{n=0}^\infty D_0(k)[\Pi(k)D_0(k)]^n=\frac{1}{k^2+m^2+\Pi_L(k)
    }P_L+\frac{1}{k^2+m^2+\Pi_T(k)}P_T\, ,\e

thus permitting a simple derivation of the mass corrections from the
self-energy $\Pi(k)$. Note that $P_L$ and $P_T$ are orthogonal projection
operators.

Now the calculation of the different contributions to the potential, listed
in eq.~(\ref{V}), will be described in some detail. This is possible due
to the extreme simplicity of the model. The explicit results are found in
appendix \ref{ahm-res}. Throughout this investigation dimensional
regularization is used. This means that the naively three dimensional parts of
the integral-sums are evaluated in $n-1=3-2\epsilon$ dimensions. This section
together with the relevant parts of the appendix describe the calculation of
the $\overline{\mbox{MS}}$-potential which will be improved by the
zero-temperature renormalization of section \ref{ahm-ren}.

The first step is the calculation of the leading order temperature corrections
to the masses. Here basically the results of ref.~\cite{BHW} have to be
supplemented with the $\epsilon$-dependent parts which produce finite
contributions to the potential due to one-loop divergences. The results, which
can be obtained using the integrals of appendix \ref{temp-int}, are given in
appendix \ref{ahm-res}.

Consider the contribution $V_R$ first. Due to the particle content of the
theory, given by Higgs particle, Goldstone boson, transverse and longitudinal
vector boson, the first term of eq.~(\ref{VR2}) reads
    \be V_{R1}=\frac{1}{2}\left\{\int dm_\vi^2I(m_\vi)+\int dm_\chi^2I(m_\chi)+
    (2-2\epsilon)\int dm_T^2I(m_T)+\int dm_L^2I(m_L)\right\}\, ,\label{V_R1}\e
with the standard temperature integral
    \be I(m)=\is dk\frac{1}{k^2+m^2}\, .\e

Similarly, the second term of eq.~(\ref{VR2}) is given by
    \be V_{R2}=-\frac{1}{2}(2-2\epsilon)\Pi_{b2,T}\,I(m_T)-\frac{1}{2}
    \Pi_{b2,L}\,I(m_L)\, .\e
Note, that this type of corrections does only exist for the vector particles
and that $\Pi_{a2}(k)=0$ in the case of the Abelian model. The sum of both
terms $V_R=V_{R1}+V_{R2}$ enters the contribution $V_4$ which can be found in
appendix \ref{ahm-res}.

The two-loop contributions of $V_z$ are of vector-scalar and of scalar-scalar
type. Figure \ref{u12l} shows the setting sun diagrams of the Abelian Higgs
model, contributing to $V_\ominus$. The labelling follows the standard model
case. As usual, dashed and wavy lines represent scalar and vector propagators
respectively. When solving the integrals, it is useful to write the propagator
as a sum of a covariant part and a longitudinal correction
    \be D(k)=D_0(k)+\left(\frac{1}{k^2+m_L^2}-\frac{1}{k^2+m^2}\right)P_L\, .\e
This is possible, because the transverse mass receives no leading order
temperature correction, due to gauge invariance.
\begin{figure}
\refstepcounter{figure}
\label{u12l}
\end{figure}
%\hspace*{1.5cm}
%\psfig{width=12.8cm,file=u12l.eps}
\\[.2cm]
{\bf Fig.\ref{u12l}} Setting sun diagrams in the Abelian Higgs model\\[.6cm]
A straightforward calculation of all the terms described above results in the
explicit formulae given in the appendix.

The necessary $\overline{\mbox{MS}}$-counterterms are generated by a
multiplicative renormalization of couplings, mass term and scalar field
    \be \la_b=Z_\la\la\quad,\quad e_b^2=Z_{e^2}e^2\quad,\quad\nu_b=Z_\nu\nu
    \quad,\quad\Phi_b^2=Z_{\vi^2}\Phi^2\, ,\label{Z}\e
where the index $b$ denotes the bare parameters. No renormalization of the
vector field is required because it does not enter the final formula for the
potential. The counterterms for the potential are now easily obtained by
expressing the bare parameters, which enter the order $g^2,\la$ part of the
potential, through the $\overline{\mbox{MS}}$-quantities, defined by
eq.~(\ref{Z}). Note that the corrections $\delta Z=Z-1$ are defined to have no
finite part. They are given explicitly in appendix \ref{ahm-res} together with
the correction to the potential which they generate.

\section{Renormalization at $T=0$}\label{ahm-ren}
To get rid of the arbitrary scale $\bar{\mu}$ the potential is rewritten in
terms of physical parameters defined at zero temperature. Such parameters are
the Higgs and vector masses and the vacuum expectation value of the physical
Higgs field $\vi_{phys}^2=Z_{\vi^2}^{-1}\vi_b^2$. Note that here, in contrast
to the $\overline{\mbox{MS}}$-definitions of the previous section, the
counterterms $\delta Z$ do have finite parts, to be specified below.

Returning for the moment to Minkowski space, the usual on-shell definitions of
field renormalization factor, Higgs mass and vector mass read
    \be Z_{\vi^2}-1=\frac{\partial}{\partial q^2}\mbox{Re}\;\Pi_\vi(q^2)
    \Big|_{q^2=m_\vi^2}\, ,\label{pm}\e\\[-1cm]
    \[ m_\vi^2+\mbox{Re}\;\Pi_\vi(m_\vi^2)=m_{\vi,phys}^2\quad,\quad
    m^2+\mbox{Re}\;\Pi(m^2)=m_{phys}^2\, .\]
They are completed by the specification of the vacuum expectation value $v$
    \be\frac{\partial V}{\partial\vi_{phys}}\Bigg|_{\vi_{phys}=v}=0\, ,
    \label{V'}\e
where the Landau gauge zero temperature effective potential to the order
$e^4,\la^2$ is given by
    \begin{eqnarray}
    V=V_{tree}+V_1&=&-\frac{\nu_b}{2}\vi^2_b+\frac{\la_b}{4}\vi^4_b-\frac{m_
    \chi^4}{64\pi^2}\left(\frac{1}{\epsilon}+\frac{3}{2}+\ln{\frac{\bar{\mu}^2}
    {m_\chi^2}}\right)\nonumber\\ \label{Vu1}\\
    &&-\frac{m_\vi^4}{64\pi^2}\left(\frac{1}{\epsilon}+
    \frac{3}{2}+\ln{\frac{\bar{\mu}^2}{m_\vi^2}}\right)-\frac{3m^4}{64\pi^2}
    \left(\frac{1}{\epsilon}+\frac{5}{6}+\ln{\frac{\bar{\mu}^2}{m^2}}
    \right).\nonumber\end{eqnarray}
The one-loop self-energies of Higgs particle and vector field in Landau gauge
are given in appendix \ref{ahm-se}. Note that in this approach, no
one-particle reducible tadpole contributions need to be considered because
they vanish due to eq.~(\ref{V'}).

Requiring the validity of the tree-level relations
    \be m^2_{\vi,phys}=2\la v^2\quad,\quad m^2_{phys}=e^2v^2\quad,\quad\nu=\la
    v^2\e
at one-loop, the above equations permit a straightforward derivation of the
necessary counterterms:
\pagebreak

    \[ \delta Z_{\vi^2}=\mbox{Re}\,\Pi_\vi'(m_\vi^2)\quad,\quad
    \delta Z_\la=-\mbox{Re}\,\Pi_\vi'(m_\vi^2)+\frac{1}{2\nu}\left(\frac{1}{v}
    V_1'(v)-\mbox{Re}\,\Pi_\vi(m_\vi^2)\right)\, ,\]\vspace{-1.5cm}

    \be{}\e\vspace{-1.4cm}

    \[\delta Z_\nu=\frac{1}{2\nu}\left(\frac{3}{v}V_1'(v)-\mbox{Re}\,\Pi_\vi
    (m_\vi^2)\right)\quad,\quad
    \delta Z_{e^2}=-\mbox{Re}\,\Pi_\vi'(m_\vi^2)-\frac{1}{m^2}\mbox{Re}\,\Pi
    (m^2) .\]
The correction to the potential can now be obtained by renormalization of its
leading order part, i.e. by inserting these counterterms into
eq.~(\ref{deltaV}). This results in a potential, explicitly independent of the
renormalization scale $\bar{\mu}$. However, in view of the principal features
of the potential considered here, the numerical effect of the performed finite
renormalization is not very important (see \fig\ref{ofsu}).

\section{Absence of a linear term}\label{lt}
In the early days of the perturbative treatment of the phase transition in the
U(1)- and SU(2)-Higgs models, the possibility of a linear $\vi$-term in the
order-$g^3$-potential has been discussed. It has however been realized that no
such term is present, if the resummation is performed correctly
\cite{BHW,DLHLL,EQZ,A}. Of course, in the present calculation the cancellation
of linear mass terms to order $g^3$ is reproduced. These terms are not
displayed in the final formulae in the appendix to make them more compact.

However, the presence of linear terms in higher orders has not been completely
clarified before. Their cancellation is claimed in ref.~\cite{A} in a
variational approach and in ref.~\cite{DLHLL} by some gauge invariance
argument which is not further specified. Therefore it is interesting to
observe that in the present result the linear $\vi$-terms cancel to order
$g^4$: Consider the explicit formulae of appendix \ref{ahm-res} at temperatures
above the barrier temperature $T_b$ defined by the vanishing of the
$\vi$-independent scalar mass term. Expanding these expressions in $\vi$ at the
point $\vi=0$, linear terms are found in $V_a$ and $V_z$, which cancel each
other exactly. This feature, formulated more precisely as
    \[ \lim_{\vi\to 0}\frac{\partial V}{\partial\vi}=0 \quad\mbox{ to all
    orders in $e$ and $\la$}\, ,\]
can be shown to survive to all orders of small couplings resummed perturbation
theory. The proof is based on an identity following from global
U(1)-symmetry:
    \be \frac{1}{\vi}\frac{\partial V}{\partial\vi}=m_\chi^2(q^2=0)\quad. \e
Obviously it suffices to demonstrate the finiteness of the self energy
$\Pi_\chi(q^2=0)$ in the limit $\vi\to 0$. Due to the positive temperature
masses of $\chi$ and $\vi$ singularities can only arise from the transverse
gauge boson propagator above the barrier temperature. Therefore diagrams of the
kind shown in \fig\ref{mchi} have to be investigated. Here the wavy lines
symbolize leading order resummed vector propagators and the blobs are full
vertices without internal vector lines, meaning the sum of all diagrams built
from scalar propagators with the correct number of external vector lines and
possibly one or two $\chi$-lines. Notice that the vector resummation affects
only the longitudinal modes, and is therefore irrelevant for the discussion of
small-$\vi$ singularities.\\[.2cm]
\begin{figure}
\refstepcounter{figure}
\label{mchi}
\end{figure}
%\hspace*{1cm}
%\psfig{width=12.8cm,file=mchi.eps}
%\\[-1.1cm]
{\bf Fig.\ref{mchi}} Higher loop self-energy corrections to $m_\chi^2$\\[.6cm]
Consider first the full vertices with external vector lines only denoted by
    \be \Gamma^{2n}_{\alpha\beta\ldots\mu\nu}(k_1,\ldots,k_{2n},\vi)\e
below. Since in the contributing diagrams all propagating particles are massive
scalars, the vertices $\Gamma$ are analytic in $k_i$.

Above the barrier temperature scalar masses are analytic in $\vi^2$. In
addition, explicit $\vi$-factors appearing at the vertices of some diagrams are
always paired, due to the structure of the unbroken theory, which has no
vertices with an odd number of scalar lines. This shows that $\Gamma$ is also
analytic in $\vi^2$.

Furthermore the three-dimensional part of $\Gamma$ with vanishing external
Matsubara frequencies satisfies
    \be \Gamma^{2n}_{\alpha\beta\ldots\mu\nu}(k_1,\ldots,k_{2n},\,\vi=0)
    \;\sim\;|\vec{k}_1|\ldots|\vec{k}_{2n}|\e
    \[\mbox{for small}\quad|\vec{k}_i|\quad\quad\mbox{and}\quad
    k^0_i=0\, ,\quad\alpha\beta\ldots\mu\nu\in\{1,2,3\}\, . \]
This follows from a gauge covariance argument, completely analogous to the zero
temperature case. Having established these properties of $\Gamma$, the
small-$\vi$ behaviour of the $\chi$-self-energy can be derived as follows:

Consider the most dangerous lowest power of $\vi$ stemming from the maximal
infrared divergence, which is obtained by setting $k^0=0$ for all transverse
vector propagators. It can be calculated by scaling the loop momenta according
to $\vec{k}\to\vec{k}\vi$. The above discussion of the pure vector vertices
$\Gamma$ shows that after this scaling they can be counted as order $\vi^2$ at
least. In the case where no scalar line connecting two external $\chi$-lines
exists (see the first diagram of \fig\ref{mchi}), the power counting in $\vi$
proceeds as follows: a factor $\vi^3$ for each of the $L$ loops, $\vi^{-2}$ for
each of the $I$ internal vector lines,  $\vi^2$ for each of the $V-2$ full
vector vertices and an explicit factor $\vi$ for each of the two vertices with
a $\chi$-line. Together this gives the minimal over all power of $\vi$
\be n_\vi=3L-2I+2(V-2)+2\quad \label{nvi}\e
for a diagram with $V$ vertices (compare the argumentation in appendix A of
ref.~\cite{BFHW}).

If there is at least one scalar line connecting the two external $\chi$-lines
(compare the second diagram of \fig\ref{mchi}), $V\!-\!1$ full vector vertices
contribute factors $\vi^2$. In this case however the last term $+2$ does not
exist because a vertex with two $\chi$-lines need not have an explicit
$\vi$-factor. Therefore eq.~(\ref{nvi}) is valid in the second case as well and
consequently in general.

Now the well-known formula $V+L-I=1$ immediately gives
\be n_\vi=L\ge 0\, , \e
or equivalently: There is no divergence for $\vi\to 0$.

If some of the vector propagators have non-zero Matsubara frequencies, the
vertices connected by those "heavy" lines may be formally fused. Now repetition
of the above argument leads again to the desired result, thus completing the
proof.

This nice feature of the Abelian model strongly supports the hope for a
reliable perturbation series in the symmetric phase. Unfortunately, due to the
3- and 4-vector vertices of the non-Abelian theory, the above argument does not
apply there.

\section{Numerical results and discussion}\label{ahm-num}

\begin{figure}[h]
\refstepcounter{figure}
\label{potu}
{\bf Fig.\ref{potu}} Different approximations of the effective potential
                     plotted at their respective critical temperatures at
                     $m_{\mbox{\scriptsize Higgs}}=38$ GeV
                     (the $e^4,\la$-potential is a result of ref.~\cite{AE})
\end{figure}
\begin{figure}[h]
\refstepcounter{figure}
\label{ofsu}
{\bf Fig.\ref{ofsu}} Dependence of the surface tension calculated from the
                     different potentials on the zero temperature Higgs mass
\end{figure}
\begin{figure}[h]
\refstepcounter{figure}
\label{ofsu34}
{\bf Fig.\ref{ofsu34}} Influence of the $e^4\vi^4$-correction on the surface
                       tension at small Higgs masses
\end{figure}

For the numerical investigation of the phase transition standard model values
are chosen for vacuum expectation value $v$ and vector boson mass $m_W$ at
zero temperature (see section \ref{su2-num}). Of course the analogy has to be
used with caution, because even at one-loop level, the three vector degrees
of freedom of the SU(2) increase the strength of the phase transition
considerably if compared to the U(1) case. Nevertheless, assuming the
standard model situation, in the following analysis the Higgs mass will be
considered as an unknown parameter.

As is well known, the potential suggesting a first order phase transition is
generated at order $g^3,\la^{3/2}$ by the combination of quadratic, cubic and
quartic $\vi$-terms, where the coefficient of the quadratic term is of the
form $T^2-T_b^2$ and therefore strongly depends on the temperature (see $V_3$
in appendix \ref{ahm-res}). Figure \ref{potu} shows the different
approximations to the potential at their respective critical temperatures and
$m_{\mbox{\scriptsize Higgs}}=38$ GeV, all of them suggesting a first order
phase transition, but of quite different strength. Calculations of order
$e^3,\la^{3/2}$ \cite{Ar,BHW}, of order $e^4,\la$ \cite{AE} and the present
$e^4,\la^2$-calculation \cite{He} are compared, showing a dramatic decrease of
the barrier height in both higher order results.

Obviously, reliability of perturbation theory has to be questioned already at
this small Higgs mass. A more detailed picture can be obtained by considering
the surface tension \cite{CL}
    \be {s}=\int_0^{\vi_+}d\vi\sqrt{2V(\vi,T_c)}\, , \label{st}\e
which may be seen as a measure of the strength of the phase transition. The
reliability of the present calculation of this quantity is of course doubtful,
because it relies essentially on the non-convex region of the potential, where
the physical interpretation is still unclear. Nevertheless ${s}$ can be used
conveniently to discuss the properties of the potential as a function of the
Higgs mass. The results are shown in \fig\ref{ofsu}.

\begin{figure}[h]
\refstepcounter{figure}
\label{lhu}
{\bf Fig.\ref{lhu}} Higgs mass dependence of the latent heat $\Delta Q$ of the
                    phase transition
\end{figure}
\begin{figure}[h]
\refstepcounter{figure}
\label{phiu}
{\bf Fig.\ref{phiu}} Position of the second minimum $\vi_+$ at the phase
                     transition in units of the critical temperature $T_c$
\end{figure}

The reasons for the difference between the $e^3,\la^{3/2}$ and the
$e^4,\la^2$-results are twofold. Consider the region of small Higgs masses
first. Here the $e^4\vi^4$-term, being a large correction to the tree-level
term $\la\vi^4/4$, is mainly responsible for the decrease of the barrier
height. This is illustrated in \fig\ref{ofsu34}, where in addition to the
consistent third and fourth order results a potential containing the third
order terms together with the $e^4\vi^4$-correction (see $V_4$ in appendix
\ref{ahm-res}) is investigated. The effect is not removed by zero temperature
renormalization, although the temperature independent constant $c_1$ forms the
largest part of the coefficient of $e^4\vi^4$. Notice that this large constant
arises from the expansion of the temperature dependent part of the one-loop
integral $I(m)$ \cite{DJ}, and is therefore absent at $T=0$.

As it can be seen from \fig\ref{ofsu34}, two-loop contributions are not too
important at small Higgs mass and perturbation theory appears to be in a
relatively good shape.

For Higgs masses above $\approx 30$ GeV another higher order effect becomes
more important: The three-dimensional pieces of two-loop temperature integrals
generate terms of the form $\vi^2\ln(m+m_{\vi,\chi})$ and the like (see
$V_a$ and $V_b$ in appendix \ref{ahm-res}), which, in spite of their numerical
smallness, influence the potential significantly. This can be understood by
recalling that at the critical temperature the leading order $\vi^2$-terms
essentially cancel and that a $\vi$-dependence in a coefficient of $\vi^2$ can
not be absorbed in a correction of $T_c$. These logarithmic terms with a
positive sign decrease the barrier height, which is clear from the shape of the
function $x^2\ln(x+\mbox{const.})$. In summary, the most infrared sensitive
contributions of the high temperature field theory introduce large corrections
and prevent the reliability of perturbation theory.

The effect of the above logarithmic terms is overestimated by the approximation
used in ref.~\cite{AE}. There, scalar masses are counted as order $\la^{1/2}$
and neglected systematically against vector masses, resulting in contributions
of the type $\vi^2\ln m$. These terms, lacking the scalar mass cutoff in the
logarithm, destroy the first order phase transition for Higgs masses above
$\approx 40$ GeV. The plot at 38 GeV of \fig\ref{potu} does already show the
arising pathology of the $e^4,\la$-potential, which can be somewhat eased
but not cured by addition of the $\la^{3/2}$ terms.

The latent heat of the phase transition is another interesting quantity to be
calculated from the effective potential :
    \be \Delta Q=T\frac{\partial}{\partial T}V(\vi_+,T)\Big|_{T_c}\, ,
    \label{lh}\e
where $\vi_+$ is the position of the asymmetric minimum of the potential $V$,
normalized to zero at the origin. This relation follows easily from the
definition of $\Delta Q$ together with the formula relating entropy and free
energy:
    \be \Delta Q =T\Delta S= T(S_{symmetric}-S_{broken})\quad,\quad
    S=-\frac{\partial V(\vi,T)}{\partial T}\, .\e
Figure \ref{lhu} shows that the change of the latent heat, introduced by the
$g^4,\la^2$-corrections, is very large, but not as dramatic as for the surface
tension. The phase transition appears to be much weaker first order at
two-loop.

The vacuum expectation value in the broken phase at $T_c$, shown in
\fig\ref{phiu}, does not reflect the dramatic change of the surface tension,
introduced by higher order corrections. This quantity, the jump of the order
parameter, seems to be the most reliable characteristic of the phase
transition, accessible in perturbation theory.

Note that the critical temperature is, here as well as in the standard model
case, very close to the uncorrected barrier temperature, defined by the
vanishing of the quadratic term in eq.~(\ref{V3-ahm}). Therefore a graphic
representation does not seem advisable.

Depending on the considered quantity and the standards to be chosen the phase
transition can be regarded as understood in principle for Higgs masses below
$30...50$ GeV. For larger masses it is likely to be much weaker first order,
although even such a qualitative description is not really well founded due
to the infrared problems. The next sections will show that reliability at
physically relevant large Higgs masses remains a problem in the non-Abelian
theory as well.

\chapter{Standard model}\label{sm}
In the following standard model discussion the technical parts will be confined
to the new features not present in the Abelian model. While the non-Abelian
character of the theory changes the numerical effect of two-loop corrections
qualitatively, fermions and the additional U(1)-symmetry are less important.
Therefore it suggests itself to concentrate on the much simpler SU(2)-Higgs
model for a detailed analysis of the two-loop effective potential.

This chapter is based on ref.~\cite{FH}.

\section{Calculation of the potential}\label{na}
To fix the notation the essential parts of the standard model Lagrangian are
given:
    \be {\cal L}={\cal L}_{Higgs}+{\cal L}_{gauge}+{\cal L}_{fermion}
    +{\cal L}_{Yukawa}. \e
Defining the covariant derivative as
    \be D_\mu=\partial_\mu+ig_1\frac{Y}{2}B_\mu+ig_2\frac{\tau^a}{2}W^a_\mu \e
the gauge part and fermionic part for $n_f$ fermion families are unambiguous.
The Higgs contribution reads
    \be {\cal L}_{Higgs}=-|D_\mu\Phi|^2+\nu|\Phi|^2-\la|\Phi|^4\quad,\quad
    \mbox{where}\quad\Phi=\frac{1}{\sqrt{2}}
    \left(\begin{array}{c}\vi_3+i\vi_4\\ \hat{\vi}+\vi_1+i\vi_2\end{array}
    \right)\label{higgs}\e
denotes the Higgs doublet and the shift has been applied according to
$\vi_1\,\to\,\hat{\vi}+\vi_1$. All fermions except the top quark are considered
to be massless. The resulting Yukawa Lagrangian is
    \be {\cal L}_{Yukawa}=-g_Y\bar{q}_L\tilde{\Phi}t_R\, ,\qquad q_L=\left(
    \begin{array}{c}t_L\\b_L\end{array}\right),\qquad\tilde{\Phi}=i\tau_2
    \Phi^*.\e
Following section \ref{res} the formal power counting rule
    \be g_1\sim g_2\sim g_Y\sim\la^{1/2} \e
is used.

The resummation of the scalar and vector degrees of freedom proceeds in analogy
to sections \ref{ds} and \ref{ahm-calc}. Explicit formulae are found in
appendix \ref{sm-res}. The vector resummation is complicated by the mixing of
the $W_3$-- and $B$--fields, characterized by the well known Weinberg angle
$\theta$ in the transverse part and by a different, temperature dependent angle
$\tilde{\theta}$ in the longitudinal part. To implement this in the evaluation
of Feynman diagrams the notation of a transverse and longitudinal propagator is
introduced:
    \be D_T=\frac{1}{k^2+m^2}P_T\quad,\quad D_L=\frac{1}{k^2+m_L^2}P_L\, .\e
Now the $W_3$-$W_3$-propagator is conveniently rewritten as
    \begin{eqnarray}D^{W_3W_3}&=&D_T^Z\cos^2\theta+D_T^\gamma\sin^2\theta+
    D_L^Z\cos^2\tilde{\theta}+D_L^\gamma\sin^2\tilde{\theta}\\
    &=&D_0^Z\cos^2\theta+D_0^\gamma\sin^2\theta+(cos^2\tilde{\theta}-\cos^2
    \theta)\left[\left(D_T+D_L^Z\right)-\left(D_T+D_L^\gamma\right)\right]\, .
    \nonumber\end{eqnarray}
Here the masses are specified by the indices $Z$ and $\gamma$.
{}From this the $B$-$B$-propagator is obtained by exchanging sine and cosine.
Similarly, the $B$-$W_3$-propagator reads
    \be D^{BW_3}=\left(D_T^\gamma-D_T^Z\right)\sin\theta\cos\theta+
    \left(D_L^\gamma-D_L^Z\right)\sin\tilde{\theta}\cos\tilde{\theta}\, .\e

Several comments are in order concerning the calculation of $V_R$. Calculating
the first term of eq.~(\ref{VR2}) the $\epsilon$-dependent part of $m_{\gamma
L}^4+m_{ZL}^4$ is needed (compare eq.~(\ref{V_R1})). It is obtained most easily
from the last line of eq.~(\ref{^4}). For the second part of $V_R$,
corresponding to the second term of eq.~(\ref{VR2}), the self energy
contribution $\Pi_{a2}(k)$, introduced in eq.~(\ref{pia}), is required. It is
nonzero for the $W$-field only and can be found in appendix \ref{sm-res}
together with the contributions of type $\Pi_{b2}$. These terms are displayed
most conveniently in the original $B$-$W_3$-basis.

$V_z$ contains two-loop contributions of vector-vector, vector-scalar and
scalar-scalar type. The setting sun diagrams are shown in
\fig\ref{sm2l},
where the labelling follows ref.~\cite{AE}, to make comparison more easy.
    \be V_\ominus=V_a+V_b+V_i+V_j+V_m+V_p .\label{Vominus}\e
Note that ghost contributions have been included in $V_m$, and that the purely
scalar diagram $V_p$ has not been considered in ref.~\cite{AE}. The calculation
of all these terms is long but straightforward. All the integrals with
complicated covariant structure can be reduced to the basic types given in
appendix \ref{temp-int}, as described in ref.~\cite{AE}.

Dropping the appropriate terms of $V$ the lower order $g^4,\la$-result, as it
is given by Arnold and Espinosa in ref.~\cite{AE}, can be derived. This is also
valid for the Abelian model discussed in the previous chapter. Another
calculation for the SU(2)-Higgs model appeared in ref.~\cite{FKRS}, where the
complete $g^4,\la^2$ result is derived from the effective three-dimensional
theory together with some corrections from ref.~\cite{AE}. The analytic results
of the actual loop-calculation are found to be in agreement with the present
analysis. However, due to a specific way to apply the renormalization group
method the numerical outcome of ref.~\cite{FKRS} differs from the results to
be presented here.\\[.2cm]
%\hspace*{1.5cm}
%\psfig{width=12.8cm,file=sm2l1.eps}\\[-1.5cm]
%\hspace*{1.5cm}
%\psfig{width=12.8cm,file=sm2l2.eps}\\[-0.7cm]
\begin{figure}
\refstepcounter{figure}
\label{sm2l}
\end{figure}
{\bf Fig.\ref{sm2l}} Setting sun diagrams for the standard model\\[.6cm]
Note that there are linear $\vi$-terms of fourth order in the couplings
present in $V_a$, $V_m$ and $V_z$. As in the Abelian model, these terms cancel
each other, thus ensuring the relation $\lim_{\vi\to 0}\partial
V/\partial\vi=0$ for all allowed temperatures. This cancellation is
essentially the same effect which leads to a vanishing third order transverse
gauge boson mass term in the symmetric phase \cite{BFHW}, as can be seen in
the contributions of diagrams fig. 6.o, 6.q, 6.t and 6.u of ref.~\cite{BFHW}.

\section{Renormalization at $T=0$}
In the Abelian Higgs model the performed zero temperature renormalization has
not affected the calculated phase transition parameters significantly.
Nevertheless, it is not possible to profit from that experience by just
setting $\bar{\mu}=1/\beta$ in the standard model case. The reason for
that is the large negative $g_Y^4\vi^4$-term which dominates over the tree
level quartic term. This leads to an $\overline{\mbox{MS}}$-potential unbounded
from below for small Higgs mass.

The zero temperature renormalization is performed in the on-shell scheme, as
described in ref.~\cite{BSH}. Slightly modifying the procedure of section
\ref{ahm-ren}, the physical parameters chosen are Higgs mass, top quark mass,
W- and Z-boson masses and the fine structure coupling $\alpha$, defined in the
Thompson limit \cite{PDG}. The physical masses and the wave function
renormalization of the Higgs field are defined in analogy to eq.~(\ref{pm}). A
multiplicative renormalization of the coupling constants, the tree level Higgs
mass square $-\nu$ and the Higgs field is performed.

The required one-loop corrections are of course well known in Feynman-'t$
\:\mbox{H}$ooft gauge \cite{BSH}. However, here they are needed in Landau
gauge. There is no problem with the correction to the electric charge
$\delta e=e_b-e$, which is gauge independent \cite{S}. This can be easily
checked explicitly using the results of ref.~\cite{DS}, where the gauge
dependence of several self-energy and vertex corrections has been calculated.
Therefore in the present calculation the formula for $\delta e$ from \cite{BSH}
is used. The logarithmic terms with the five light quark masses are treated in
the way described in ref.~\cite{Ho}, with data from ref.~\cite{J}, resulting
in the vacuum polarization contribution
    \be \mbox{Re}\,\hat{\Pi}^{\gamma(5)}_{\mbox{\footnotesize had}}(M_Z^2)=
    \mbox{Re}\,\Pi^{\gamma(5)}_{\mbox{\footnotesize had}}(M_Z^2)-\Pi^{
    \gamma(5)}_{\mbox{\footnotesize had}}(0)=-0.0282\pm 0.0009\, .\e
The dependence of the one-loop self energy corrections on the gauge parameters
has been calculated in ref.~\cite{DS} for gauge bosons. Therefore the
corrections in Landau gauge, needed here, can be taken from \cite{DS,MS}. The
self energy corrections for the physical Higgs boson and the top quark can be
easily calculated in Landau gauge. The results are displayed in appendix
\ref{sm-se}.

Defining $v$ as in the Abelian model by eq.~(\ref{V'}), no one-particle
reducible tadpole diagrams need to be added to the self energies. The zero
temperature one-loop effective potential $V_{tree}+V_1$ , required for that, is
easily obtained from eq.~(\ref{Vu1}) by including the additional degrees of
freedom. The renormalized couplings and the renormalized Higgs mass square are
defined by
    \be c=\frac{M_W}{M_Z}\quad,\quad cg_1=sg_2=e\quad,\quad \la=\frac{M_H^2
    g_2^2}{8M_W^2}\quad,\quad M_H^2=2\nu\, ,\e
where $s=\sin\theta_W$ and $c=\cos\theta_W$. Now the counterterms follow easily
from the definitions of the physical parameters:\\[0cm]
    \[\!\!\!\!\!\!\!\!\!\!
    \delta Z_{g_1^2}=2\frac{\delta e}{e}-\frac{\Pi_Z}{M_Z^2}+\frac{\Pi_W}
    {M_W^2}\, ,\quad\delta Z_{g_2^2}=2\frac{\delta e}{e}+\frac{c^2}{s^2}
    \frac{\Pi_Z}{M_Z^2}-\frac{c^2}{s^2}\frac{\Pi_W}{M_W^2}\, ,\quad\delta
    Z_{g_Y}=\frac{\delta m_t}{m_t}-\frac{1}{2}\Pi_\vi'\]\vspace{-1.5cm}

    \be{}\label{delZ}\e\vspace{-1.1cm}

    \[\!\!\!\!\!\!\!\!\!\!
    \delta Z_\la=2\frac{\delta e}{e}+\frac{c^2}{s^2}\frac{\Pi_Z}{M_Z^2}+
    \frac{s^2-c^2}{s^2}\frac{\Pi_W}{M_W^2}-\frac{\Pi_\vi}{M_H^2}+\frac{g_2V_1'}
    {2M_WM_H^2}\, ,\quad\delta Z_\nu=-\frac{\Pi_\vi}{M_H^2}+\frac{3g_2V_1'}
    {2M_WM_H^2}\,  \, .\]\\[-.2cm]
Here $\Pi$ and $\Pi'$ stand for the real parts of the self-energies and their
derivatives at the on-shell point.

The renormalized, $\bar{\mu}$-independent potential can now be obtained by
applying eqs.~(\ref{delZ}) and the field renormalization $\delta Z_{\vi^2}=
\Pi_\vi'$ to the leading order contribution, given by the first line of
eq.~(\ref{V3-sm}).

Clearly, the resulting formula for the potential is too long to be given
explicitly. However, it seems worthwhile to give the numerically most
important parts of the corrections to enable a simplified usage of the
analytic result in the appendix. As it has already been mentioned, the main
contributions come from the $g_Y^4$-corrections to parameters of order $\la$
(see also \cite{AE}):
    \be \delta\la=\frac{3g_Y^4}{8\pi^2}\ln\frac{m_t}{\bar{\mu}}\quad,\quad
    \delta\nu=\frac{3g_Y^4v^2}{16\pi^2}.\label{4}\e
Introducing these corrections in all terms in the potential contributing to
order $\la$ and using standard model tree level relations to calculate the
couplings one  obtains a result which is `partially renormalized at zero
temperature'. The corresponding correction to the
$\overline{\mbox{MS}}$-potential reads
    \be\delta V=\frac{\vi^2}{2}\left(-\delta\nu+\frac{1}{2\beta^2}\delta\la
    \right)+\frac{\delta\la}{4}\vi^4\, .\label{10}\e
As it will be seen later (section \ref{sm-num}), the numerical effect of this
drastic simplification is not too severe.

\section{Numerical results and discussion}

\begin{figure}[h]
\refstepcounter{figure}
\label{ofs1}
{\bf Fig.\ref{ofs1}} The surface tensions calculated from the different
                     potentials as functions of the zero temperature Higgs mass
\end{figure}

\begin{figure}[h]
\refstepcounter{figure}
\label{ofs3}
{\bf Fig.\ref{ofs3}} Influence of the most infrared sensitive contributions on
                     the surface tension as a function of the Higgs mass
\end{figure}

\begin{figure}[h]
\refstepcounter{figure}
\label{lhsu2}
{\bf Fig.\ref{lhsu2}} Higgs mass dependence of the latent heat $\Delta Q$ of
                      the phase transition
\end{figure}

\subsection{SU(2)-Higgs model}\label{su2-num}
To obtain an understanding of the qualitative effects of higher order
corrections the pure SU(2)-Higgs model is studied first. In this subsection the
additional U(1)-symmetry and the effect of fermions are neglected. A discussion
of this simplified version is also useful in view of lattice investigations,
which deal with the pure SU(2)-Higgs model presently and probably also in the
near future.

The relevant potential can be
easily derived from the formulae given in the appendix \ref{sm-res} by
performing the limit $g_1,g_Y\to 0$ and setting the number of families $n_f$ to
zero. This results in the potential of appendix \ref{su2-res}, which is the
basis for the numerical investigation of the present subsection.

Standard model values for W-mass and vacuum expectation value $v$ are used,
unless stated otherwise : $M_W=80.22$ GeV and $v=251.78$ GeV. The parameter
$\bar{\mu}$ of dimensional regularization is set to $T=1/\beta$. This can be
justified by the small dependence on the renormalization procedure. The
differences between the results obtained in this scheme and in a scheme with
on-shell $T=0$ renormalization are relatively small. This phenomenon is
observed in the Abelian Higgs model as well.

In analogy to the investigation performed in section \ref{ahm-num} the
potentials from calculations to order $g^3,\la^{3/2}$ \cite{Ca,BFHW}, to order
$g^4,\la$ \cite{AE} and to order $g^4,\la^2$ \cite{FH} are compared. All
suggest a first order phase transition in a wide Higgs mass range. The form of
the potentials at the critical temperature is the standard one (see
\fig\ref{potu}) and will not be shown here again. However, comparing the
barrier height in the different approximations the picture differs
significantly from the Abelian case in the region of large Higgs masses: Both
$g^4,\la$- and $g^4,\la^2$-potential suggest a much stronger first order phase
transition than the lowest order result. The pathological behaviour of the
$e^4,\la$-potential of the Abelian model does not arise.

In the following, the surface tension (see eq.~(\ref{st})) will be used to
illustrate the features of the potentials at different Higgs masses. Figure
\ref{ofs1} shows the similarity of the situations in the Abelian and the
non-Abelian case in the small Higgs mass region. The $g^4\vi^4$-term from the
one-loop vector contribution is responsible for the strong decrease of the
barrier height in the higher order results.

\begin{figure}[h]
\refstepcounter{figure}
\label{phisu2}
{\bf Fig.\ref{phisu2}} Position of the second minimum $\vi_+$ in units of the
                     critical Temperature $T_c$
\end{figure}

\begin{figure}[h]
\refstepcounter{figure}
\label{ofs6}
{\bf Fig.\ref{ofs6}} Surface tensions from the third and fourth order
                     potentials for a model with\linebreak $M_W=20$ GeV in the
                     Higgs mass range where the values differ by a factor of 2
                     at most
\end{figure}

At larger Higgs masses, approximately above 40 GeV, the infrared two-loop
contributions become more important. Their effect is however quite different
from the Abelian case. Compare the results of order $g^3,\la^{3/2}$ and
$g^4,\la^2$ first. The increase in the strength of the phase transition,
studied already in ref.~\cite{BD}, can be traced back to the infrared features
of a non-Abelian gauge theory. The crucial contribution is the one coming from
the non-Abelian setting sun diagram (\fig\ref{sm2l}.m). It produces
contributions to the potential of type $\varphi^2\ln(\beta m_W)$ with a
negative sign. Recall, that the logarithmic terms from diagrams \ref{sm2l}.a
and \ref{sm2l}.b (or \ref{u12l}.a, \ref{u12l}.b), discussed in the Abelian
case, have a positive sign. Notice also, that the new terms are non-analytic at
$\vi=0$. The reason why these kind of corrections affect the form of the
potential strongly has already been discussed in section \ref{ahm-num}. To
demonstrate its importance the $\varphi^2\ln(\beta m_W)$-term of $V_m$ has been
deleted by hand. The corresponding surface tension is shown in \fig\ref{ofs3}
(long-dashed line).

The comparison of the potentials to order $g^4,\la$ \cite{AE} and
$g^4,\la^2$ shows a picture very similar to the Abelian Higgs model.
Neglecting scalar masses in the logarithms results in spurious
$\varphi^2\ln(\beta m_W)$-terms, which reduce the surface tension. Another
important contribution, less significant in the Abelian case, is the one
proportional to $g^2(m_1+3m_2)m_{WL}$ from $V_z$. This term comes from
scalar-vector diagrams of type of \fig\ref{s2l}.b and it was neglected in
ref.~\cite{AE}. On the relevant scale ($\varphi<T$) it produces a very steep
behaviour of the potential, again increasing the surface tension. The observed
difference between the result of ref.~\cite{AE} and the complete $g^4,\la^2$
calculation presented here is mostly due to these two effects, together with
the well known influence of the cubic scalar mass contributions from $V_3$.
However, in sharp contrast to the Abelian case, both the $g^4,\la$- and the
$g^4,\la^2$-calculation suggest a much stronger first order phase transition
than the lowest order result.

\begin{figure}
\refstepcounter{figure}
\label{ofs4}
{\bf Fig.\ref{ofs4}} The surface tensions obtained from the standard model
                     effective potentials as functions of the zero temperature
                     Higgs mass with $m_{\mbox{\scriptsize top}}=170$ GeV
\end{figure}
\begin{figure}
\refstepcounter{figure}
\label{phism}
{\bf Fig.\ref{phism}} Position of the second minimum $\vi_+$ in units of the
                     critical temperature $T_c$ in the case of the standard
                     model with $m_{\mbox{\scriptsize top}}=170$ GeV
\end{figure}

Another interesting effect of higher order $\la$-corrections is the complete
breakdown of the phase transition at a Higgs mass of about 100 GeV, where the
surface tension is very large. In this region the above mentioned term,
proportional to $g^2(m_1+3m_2)m_{WL}$, becomes important. For a temperature
close to the uncorrected barrier temperature $T_b$, at which the scalar masses
vanish for $\vi=0$, it produces an almost linear behaviour in the small $\vi$
region. This results in a potential for which at $T=T_b$ the asymmetric minimum
is not a global minimum but only a local one. Note that $T_b$ is the lowest
temperature accessible in this calculation. In other words, the temperature
region in which the phase transition occurs can not be described by the given
method, due to infrared problems.

In order to illustrate the possible effects of the unknown infrared behaviour
of the transverse vector propagator, the dependence of the surface tension on
the magnetic mass can be studied. A magnetic mass of order $g^2T/3\pi$ is
motivated by the solution of gap equations \cite{BFHW} and supported by the
numerical investigation of gauge invariant gap equations with resummed vertices
\cite{BP}. Following the approach of ref.~\cite{BFHW} the transverse vector
mass takes the form
    \be m_W^2=\left(\frac{g\vi}{2}\right)^2+\left(\frac{\gamma g^2}{3\pi\beta}
    \right)^2\, , \e
where $\gamma$ is some unknown parameter. One can introduce this redefined
transverse mass in the most influential infrared contributions, i.e. in the
$m_W^3$- and in the $\vi^2\ln\beta m_W$-terms. Figure \ref{ofs3} shows the
results obtained for $\gamma=$ 0, 2 and 4, supporting the qualitative behaviour
found in ref.~\cite{BFHW}. The main difference is due to the fact that the
higher order result suggests a stronger first order phase transition, thus for
a given $m_{\mbox{\scriptsize Higgs}}$ a larger magnetic mass is necessary to
change the phase transition to second order.

A complete fourth order calculation of the surface tension has to include the
wave function correction term $Z_\vi(\vi^2,T)$ calculated in ref.~\cite{BBFH}.
Using the results of ref.~\cite{BBFH} the surface tension
    \be {s}=\int_0^{\vi_+}d\vi\sqrt{2\,(1+Z_\vi(\vi^2,T))\,V(\vi,T_c)}\, ,\e
has been determined for Higgs masses between 25-95 GeV. The numerical effect of
this $Z$-factor is very small, only $1\%-4\%$. This is due to the smallness of
the potential in the only region where $Z_\vi$ is significant, i.e. at small
$\vi$.

The latent heat $\Delta Q$ (see eq.~(\ref{lh}))is plotted in \fig\ref{lhsu2} as
a function of the Higgs mass. In contrast to the Abelian model, here the higher
order results show an almost linear increase of the latent heat with the Higgs
mass. This somewhat surprising behaviour can be understood by observing that
for those potentials neither the position of the degenerate minimum nor the
height of the barrier change significantly with increasing Higgs mass (see
\fig\ref{ofs1}). Therefore $\partial V/\partial T$ does not change dramatically
over a wide Higgs mass range. On the other hand the critical temperature
increases with growing $m_{\mbox{\scriptsize Higgs}}$.

As in the Abelian case the quantity $\vi_+/T_c$, shown in \fig\ref{phisu2} as a
function of the Higgs mass, is least affected by higher order corrections.

Now the question arises whether a good convergence of the perturbation series,
which can not be claimed in the whole range of $\la$ for a realistic gauge
coupling $g=0.64$, could be present in the region of much smaller gauge
coupling constants. This seems indeed to be the case, as can be seen in fig.
\ref{ofs6}, where the surface tensions of order $g^3,\lambda^{3/2}$ and
$g^4,\lambda^2$ are plotted for a model with a vector mass of 20 GeV, i.e.
$g=0.16$. In the used Higgs mass range the two results for ${s}$ differ by a
factor of two at most. The relative size of this range, i.e. the ratio of the
minimal and maximal values of the Higgs mass, is 4, which is twice as large as
the corresponding range for the model with $M_W=80$ GeV.

The phase transition parameters calculated in this section can be compared to
new lattice data available at a low Higgs mass point
($m_{\mbox{\scriptsize Higgs}}\approx 18$~GeV) and a high point
($m_{\mbox{\scriptsize Higgs}}\approx 49$~GeV) \cite{L1,L2}. These data include
critical temperature, jump of the order parameter $\Phi^\dagger\Phi$, latent
heat and surface tension at both Higgs mass values. The $g^4,\la^2$-results are
in good quantitative agreement with lattice data (explicit numbers will appear
soon \cite{BFH1}). This is highly non-trivial, since quantities change by
large factors between low and high point. An exception is formed by the surface
tension at the high point, which is larger by a factor of $\sim2.5$ in
perturbation theory. All other parameters agree between the perturbative and
the lattice calculation with deviations compatible with the observed scaling
violation on the lattice and the uncertainties of perturbation theory.
Note, that the simulations have been performed with a maximum number of three
lattice points in time-like direction, thus the results may change somewhat on
larger lattices. To obtain the good agreement quoted above the renormalization
effect of the vector boson on the very light Higgs particle has to be taken
into account at the low point \cite{BFH1}.

\subsection{Complete standard model}\label{sm-num}
In the case of the full standard model the qualitative behaviour of the
potential is essentially the same as for the SU(2)-Higgs model. The main
difference is a decrease of the surface tension by a factor $\sim 4$. This can
be traced back to the large top quark mass which influences the potential by
lowering the barrier temperature (see \ref{V3-sm}). The additional
U(1)-symmetry and the light fermions are less important. Also the
characteristic points of the surface tension plot of \fig\ref{ofs1} are shifted
to higher values of the Higgs mass. Figure \ref{ofs4} shows the surface tension
as a function of $m_{\mbox{\scriptsize Higgs}}$. The complete breakdown of the
$g^4,\la^2$ calculation, observed at $m_{\mbox{\scriptsize Higgs}}\approx 100$
GeV for the pure SU(2) case, occurs at $m_{\mbox{\scriptsize Higgs}}\approx
200$ GeV in the full model. These quantitative differences do not change the
qualitative features of the potential, thus the discussion given in the
previous section does also apply to the standard model. The difference between
the fully renormalized potential and the partially renormalized potential (see
eqs.~(\ref{4}),(\ref{10})) is not too severe in view of the huge uncertainties
still present in the perturbative approach. Again, the position of the second
minimum at the critical temperature given in \fig\ref{phism}, does not depend
as strongly on the order of the calculation as the height of the barrier.
Unfortunately, the region $m_{\mbox{\scriptsize Higgs}}\approx 40$~GeV, in
which the reliability of the perturbative approach is the best and $\vi_+/T_c
\approx 1$, is well below the experimental Higgs mass bound.

\chapter{Gauge invariant approach}\label{gi}
In this chapter the gauge invariant treatment of the electroweak phase
transition is presented following ref.~\cite{BFH}. The main new point is the
introduction of a gauge invariant source term \cite{Lu}
    \be Z=\exp(-\beta\Omega W)=\mbox{\bf tr}\exp\left[-\beta(\bm{H}+\int_V
    J\,\bm{\Phi^\dagger\Phi})\right] \e
and the subsequent definition of an effective potential by the Legendre
transformation
    \be V(\sigma,T)=W(J,T)-J\frac{\sigma}{2}\qquad,\qquad\frac{\sigma}{2}=
    \frac{\partial W(J,T)}{\partial J}\, ,\label{Vsigma}\e
which is performed perturbatively. The new variable $\sigma/2=<\bm{\Phi^\dagger
\Phi}>$ describes the thermal expectation value of a gauge invariant quantity.

This approach is perfectly well suited for comparison with lattice
investigations, which usually proceed without gauge fixing and consider the
expectation value of the operator $\bm{\Phi^\dagger\Phi}$. Note also, that
effective actions for composite operators have been defined previously, for
example in ref.~\cite{CJT}, and that a formulation based on a field linearly
related to $\phi^2$ has been given in ref.~\cite{Law}.

\section[One-loop calculation of $W(J)$ for the SU(2)-Higgs model]{One-loop
calculation of $W(J)$ for the \hspace{4cm}\mbox{} SU(2)-Higgs model}
\label{WJ}
The analysis is restricted to the pure SU(2)-Higgs model, which is sufficient
to illustrate the main features of this method at the one-loop level. Note that
a comparison with the Abelian model makes no sense at one-loop level since
qualitative differences start to appear only at the next loop order. Using the
Lagrangian ${\cal L}$ from section \ref{na}, where the appropriate limits have
been taken (see section \ref{su2-num}), the thermodynamic potential $W$ can be
calculated from the path integral
    \be \exp(-\beta\Omega W)=\int D\Phi D\Phi^\dagger DW_\mu\exp\left[
    \int_0^\beta d\tau\int d^3\vec{x}({\cal L}-J\Phi^\dagger\Phi)\right]\, .\e
To evaluate this integral the extremum of the static, $\Phi$-dependent part of
the action with source term, characterized by $V_{tree}$, has to be determined:
    \be {\cal L}^J={\cal L}-J\Phi^\dagger\Phi=-V_{tree}+\cdots\, ,\e
    \be V_{tree}=(-\nu+J+\alpha_0 T^2)\,\Phi^\dagger\Phi
    +\la(\Phi^\dagger\Phi)^2\, .\label{Vtree}\e
Here, following the philosophy described in section \ref{ctm}, a thermal
counterterm for the scalar field has been introduced. Its leading and next to
leading order contributions read
    \be \alpha_0=\alpha_{01}+\alpha_{02}=\left(\frac{1}{2}\la+\frac{3}{16}g^2
    \right)-\frac{3g^3}{16\pi}\sqrt{\frac{5}{6}}\, .\e
The $g^3$-term, generated by the one-loop self-energy contribution of
the longitudinal vector boson (see ref.~\cite{BFHW}, eqs. (41),(42)), is taken
into account for reasons to be explained below.

Two different regimes have to be distinguished for the calculation of $W(J)$.
Consider the case with $J<\nu-\alpha_0T^2$ first. In this situation $V_{tree}$
develops a non-trivial minimum at $\vi_1=\hat{\vi}(J)$ (see eq.~(\ref{higgs})
for the conventions), defined by
    \be \la\hat{\vi}^2=M^2=\nu-J-\alpha_0T^2\, .\label{Mb}\e
Performing the shift $\vi_1\,\to\,\hat{\vi}+\vi_1$ a tree-level contribution
    \be W_{tree}(J)=\frac{1}{2}(-\nu+J)\hat{\vi}^2+\frac{\la}{4}\hat{\vi}^4\,
    ,\e
which is independent of the resummation, is generated. The one-loop
contribution in Lorentz gauge (cf. ref.~\cite{DJ}) with gauge fixing parameter
$\alpha$ reads
    \[ W_1(J) =\frac{1}{2}\is dk\Bigg[9\ln(k^2+m^2)+\ln(k^2+m_1^2)+3\ln(k^4
    +k^2m_2^2+\alpha m^2m_2^2)-6\ln k^2\Bigg]\]
    \be+\frac{1}{2}\int\frac{d^3\vec{k}}{{(2\pi)^3}\beta}\Bigg[3\ln(\vec{k}^2+
    m_L^2)-3\ln(\vec{k}^2+m^2)\Bigg]\, .\label{W1}\e
Here Higgs mass, Goldstone mass, vector mass and longitudinal vector mass are
given by
    \be m_1^2=2M^2\quad,\quad m_2^2=0\quad,\quad m^2=\frac{1}{4}g^2\hat{\vi}^2
    \quad,\quad m_L^2=\frac{1}{4}g^2\hat{\vi}^2+\frac{5}{6}g^2T^2\, .\e
Note that only the zero Matsubara frequency mode of the longitudinal vector
degree of freedom has been resummed. This simplifies the calculation in
Lorentz gauge and does not change the result up to order $g^3$ extracted from
the one-loop formula.

If, on the other hand, $J>\nu-\alpha_0T^2$, no shift of the field is necessary
and $W(J)$ has no tree-level contribution. The one-loop term can be easily
obtained from eq.~(\ref{W1}) by setting
    \be m_1=m_2=M\qquad\mbox{with}\qquad M^2=-\nu+J+\alpha_0T^2\label{Ms}\e
and replacing transverse and longitudinal vector masses by zero and
$m_{L,0}=\sqrt{5/6}gT$ respectively.

In both the symmetric and the broken phase eq.~(\ref{W1}) gives explicitly
gauge independent results since the product $m_2m$ is always zero. Working in
$R_\xi$-gauge the same answer is obtained. The integrals are easily performed
using appendix \ref{temp-int} and terms of fourth and higher order in the
masses are neglected together with constant terms common to both phases.

The one-loop thermodynamic potential $W(J)$ in broken and symmetric phase is
given by
    \be W(J)=W_b(J)\Theta(\nu-J-\alpha_0T^2)+W_s(J)\Theta(-\nu+J+\alpha_0T^2)
    \, ,\e
where
    \begin{eqnarray}W_b(J)&=&-\frac{T^2}{6}M^2-\frac{1}{4\la}M^4\label{Wb}\\
    &&-\frac{T}{12\pi}\left[(2M^2)^{3/2}+6m^3+3m_L^3\right]-\frac{\vi^2}{2}
    \alpha_{02}T^2\, ,\nonumber\\ \nonumber\\
    W_s(J)&=&\frac{T^2}{6}M^2-\frac{T}{12\pi}\left[4M^3+3m_{L,0}^3\right]
    \label{Ws}\end{eqnarray}
and $M^2$ is given by eqs.~(\ref{Mb}) and (\ref{Ms}) respectively.

\section{Gauge invariant effective potential}\label{gip}
The one-loop gauge invariant effective potential can now be obtained by a
perturbative Legendre transformation according to eq.~(\ref{Vsigma}). In
the usual approach, based on the order parameter $\vi$, the perturbatively
defined effective potential is the sum of the one-particle irreducible vacuum
graphs of the shifted theory. However, no such interpretation is known for the
gauge invariant potential $V(\sigma)$. Therefore, after writing $W(J)$ and $J$
as a sum of contributions of increasing order
    \be W(J)=W_0(J)+W_1(J)+W_2(J)+\cdots\qquad,\qquad J=J_0+J_1+J_2+\cdots\,
    ,\e
the definition (\ref{Vsigma}) has to be implemented order by order in
perturbation theory \cite{BFH,La1}. The first terms of the resulting
potential are given by
    \be V(\sigma)=\Bigg[W_0(J_0)-\frac{\sigma}{2}J_0\Bigg]+\Bigg[W_1(J_0)\Bigg]
    +\Bigg[W_2(J_0)-\frac{1}{2}\left(\frac{\partial W_1(J)}{\partial J}\bigg|_{
    J_0}\right)^2\Bigg/\frac{\partial^2W_0(J)}{\partial J^2}\bigg|_{J_0}\Bigg]
    +\cdots\, ,\label{VfW}\e
where $J_0$ is a function of $\sigma$ defined by the leading order relation
    \be \frac{\partial W_0(J)}{\partial J}\bigg|_{J_0}=\frac{\sigma}{2}\, .\e

Consider the phase with broken symmetry first. Here $W(J)$ has to be decomposed
into a leading term $W_0$ (order $g^2,\la$) and a next to leading term $W_1$
(order $g^3,\la^{3/2}$), defined by the first and second line of eq.~(\ref{Wb})
respectively.

In the symmetric phase, where no tree-level term exists, the Legendre
transformation can be performed exactly at one-loop order. Neglecting constant
terms common to both phases the resulting potential is given by
    \be V(\sigma)=V_s(\sigma')\Theta(\sigma')+V_b(\sigma')\Theta(-\sigma')\quad
    ,\quad \sigma'=\sigma-\frac{T^2}{3}\, ,\e
where
    \begin{eqnarray}V_b(\sigma')&=&\frac{1}{2}(\alpha_{01}T^2-\nu)\sigma'+
    \frac{\la}{4}\sigma'^{\,2}\label{Vb}\\
    &&-\frac{T}{12\pi}\left[(2\la\sigma')^{3/2}+6\left(\frac{1}{4}g^2\sigma'
    \right)^{3/2}+3\left(\frac{1}{4}g^2\sigma'+\frac{5}{6}g^2T^2\right)^{3/2}
    \right]\, ,\nonumber\\ \nonumber\\
    V_s(\sigma')&=&\frac{1}{2}(\alpha_0T^2-\nu)\sigma'-\frac{\pi^2}{6}\frac{
    \sigma'^{\,3}}{T^2}-\frac{T}{4\pi}\left(\frac{5}{6}g^2T^2\right)^{3/2}
    \label{Vs}\, .\end{eqnarray}
Here the new variable $\sigma'$ has been introduced to separate the shift of
the field variable generated by the interaction from the basic, model
independent thermal expectation value $T^2/3$ \cite{KL2}. Note the similarity
of $V_b$ to the order-$g^3,\la^{3/2}$ Landau gauge result, eq.~(\ref{V3-su2}).
The new values of the scalar masses, namely zero for the Goldstone boson mass
and $2\la\sigma'$ for the Higgs mass square, form the main difference of $V_b$
and $V_3$.

\begin{figure}[h]
\refstepcounter{figure}
\label{potsgi}
{\bf Fig.\ref{potsgi}} Gauge invariant effective potential at critical
                       temperature, $m_{\mbox{\scriptsize Higgs}}$=70GeV
                       \\[.5cm]
\refstepcounter{figure}
\label{pots}
{\bf Fig.\ref{pots}} Effective potential in Landau gauge at critical
                     temperature, $m_{\mbox{\scriptsize Higgs}}$=70 GeV
\end{figure}

The gauge invariant potential defines a critical temperature at which its two
minima, one in the symmetric and one in the broken phase, are degenerate. This
is illustrated in \fig\ref{potsgi}. For comparison, the Landau gauge potential
to order $g^3,\la^{3/2}$ is plotted at its critical temperature in
\fig\ref{pots}. The interaction induced shift $\sigma'$ seems to play a role
similar to the squared field expectation value $\vi^2$ of the usual approach
(compare the discussion in section \ref{comp}).

The main difference of the gauge invariant and the Landau gauge potential lies
in the description of the symmetric phase. Choosing the expectation value of
the basic field operator as order parameter the symmetric phase corresponds to
exactly one point, where this parameter must vanish. In contrast to this, the
gauge invariant coupling of the source permits a description of the symmetric
phase by a non-trivial minimum of the free energy as a function of the field
square expectation value. This expectation value can be made smaller than its
value at the minimum by turning the source, i.e. the mass term, on. Therefore
the gauge invariant potential exhibits a very steep behaviour left from the
symmetric minimum, but no definite smallest value of $\sigma'$.

\begin{figure}[h]
\refstepcounter{figure}
\label{lhgi}
{\bf Fig.\ref{lhgi}} Latent heat as a function of the zero temperature Higgs
                     mass, calculated from the potential in Landau gauge and
                     from the gauge invariant potential
\end{figure}
\begin{figure}[h]
\refstepcounter{figure}
\label{phigi}
{\bf Fig.\ref{phigi}} Shift of the Higgs field as a function of the Higgs mass,
                      calculated from the potential in Landau gauge and from
                      the gauge invariant potential
\end{figure}

Note, that the gauge invariant potential is continuous together with its first
derivative at the matching point $\sigma'=0$. This is due to the difference of
the first terms of eqs.~(\ref{Vb}) and (\ref{Vs}) which compensates the
contribution of the last term of eq.~(\ref{Vb}) to the derivative at
$\sigma'=0$. The additional resummation up to order $g^3$ performed in section
\ref{WJ} is responsible for that matching. However, the same result for $V(
\sigma)$ can also be obtained with leading order resummation only. In that case
the contribution
    \be \frac{3g^2T^2}{32\pi^2}Mm_{L,0}=\frac{1}{2}\alpha_{02}T^2\sigma'\e
from the scalar-vector diagram of the type of \fig\ref{s2l}.b appears in the
symmetric phase, thus rescuing the above matching of the first derivatives.

At small Higgs masses the symmetric minimum of the gauge invariant potential is
not important numerically. Therefore in this region the form of the potential
at the critical temperature and the derived phase transition parameters are
very close to the Landau gauge results. However, already at
$m_{\mbox{\scriptsize Higgs}}=70$ GeV the symmetric minimum leads to an
increase of the barrier height by a factor of $\approx 2$ (compare
figs.\hspace{.1cm}\ref{potsgi} and \ref{pots}). This effect becomes even more
important for larger Higgs masses. The latent heat plotted in \fig\ref{lhgi}
does also show a stronger first order phase transition at large Higgs mass
values suggested by the new approach. Note that a similar behaviour of the
latent heat as a function of the Higgs mass has been obtained from the two-loop
results in Landau gauge (see \fig\ref{lhsu2}). The jump of the order parameter
shown in \fig\ref{phigi} is not seriously affected by the new approach. This
justifies the conjecture of section \ref{su2-num} that this parameter is
reliably determined by the Landau gauge calculation.

Due to the separate treatment of the symmetric and the broken phase in the
gauge invariant approach, it is not clear at present how the surface tension
can be calculated. However, due to the strong increase of the barrier height
at large Higgs masses in the new approach significant changes with respect to
the Landau gauge results are expected.

\section{Problems at higher orders}
It is, in principle, straightforward to extend the calculation of the previous
sections to the order $g^4,\la^2$, as it has been done in the conventional
Landau gauge calculation of chapters \ref{ahm} and \ref{sm}. Nevertheless,
the concrete realization of this project is hampered by some problems to be
discussed in the sequel.

The first step is to supply the leading order results for $W(J)$, given by eqs.
(\ref{Wb}) and (\ref{Ws}), with higher order corrections. It is advantageous
to calculate in Landau gauge, although of course the final result should be
gauge independent by definition. The necessary corrections include the next
term of the high temperature expansion of the one-loop integrals and the
leading contributions of the two-loop graphs. The later ones, which form the
main part of the calculation, can be obtained using the formulae of appendix
\ref{su2-res} and changing the masses appropriately. Note that in the broken
phase, one-particle reducible two-loop graphs have to be considered.

To make $W(J)$ finite, the parameters of the Lagrangian are renormalized
multiplicatively in the conventional way, using e.g. the SU(2)-limit of
eqs.~(\ref{ct-sm}). Note that the multiplicative mass renormalization has to
be applied to both $-\nu$ and $J$. However, a finite result is obtained only
after the additional counterterm
    \be W_{c.t.}=\frac{1}{16\pi^2\epsilon}(J-\nu)^2\e
has been added to $W(J)$ in both the symmetric and the broken phase calculation
(compare ref.~\cite{Z}). It is claimed that a possible finite part of this
counterterm, common to both phases, does not affect the physical parameters
derived from the calculation. Firstly, constant and linear term in $J$ do only
shift the Legendre transformed function along vertical and horizontal axis
respectively. Applying such a shift to both phases does not change the
description of the phase transition. Secondly, the quadratic term in $J$ does
not affect the minimum of the Legendre transformed function, since at the
minimum $J=0$ and the quadratic correction vanishes together with its first
derivative \cite{Law}. To see this more explicitly introduce the corrected
function $\tilde{W}(J)$, defined by
    \be \tilde{W}(J)=W(J)+CJ^2.\e
The new position of the minimum of the potential is given by
    \be \frac{\tilde{\sigma}_{min}}{2}=\frac{\partial \tilde{W}}{\partial J}
    \Bigg|_{J=0}=\frac{\partial W}{\partial J}\Bigg|_{J=0}=\frac{\sigma_{min}}
    {2}\, ,\e
and analogously
    \be \tilde{V}_{min}=\tilde{W}(0)=W(0)=V_{min}\, .\e
Performing the Legendre transformation perturbatively a higher order dependence
of the result on the finite part of $W_{c.t.}$ is nevertheless present.

Now the obvious way to proceed is to calculate $V(\sigma)$ using
eq.~(\ref{VfW}). In the broken phase $W_2$ is given by the order $g^4,\la^2$
corrections and the first three terms of the perturbative expansion of $V$ can
be calculated. In the symmetric phase, where no tree-level term exists, these
two-loop corrections define $W_1$ and only the first two terms of
eq.~(\ref{VfW}) are to be used.

It has been checked that the complete potential is continuous at the matching
point $\sigma'=0$. However, its first derivative is logarithmically divergent
at this point. Since all physical information is extracted from the minima of
the potential this pathology should, in principle, have no importance.

The convincing form of the potential illustrated in \fig\ref{potsgi} has been
obtained due to the additional resummation of the scalar mass. More precisely,
the scalar mass has been resummed to the same order to which the whole
calculation was performed. Therefore it appears necessary to extend the
resummation to the order $g^4,\la^2$ in this section. Unfortunately, this can
not be done in a straightforward manner, because the two-loop scalar
self-energy is divergent at zero momentum. Another possible way to organize the
higher order resummation would be to demand the continuity of the first
derivative of the potential. This fails due to the same infrared divergences.
However, it has been seen numerically that such a higher order resummation
can change the physical picture of the phase transition significantly.

It is not known at present, how the above problems, which are essentially
connected with the infrared divergences of a massless three-dimensional field
theory, can be resolved.

\section[On the relation to the conventional effective potential]{On the
relation to the conventional effective\hspace*{1.5cm}\mbox{} potential}
\label{comp}
The gauge invariant thermodynamic potential $W(J)$, introduced in the beginning
of this chapter, has a simple relation to the effective potential, as it can be
calculated in any fixed gauge \cite{Z1}:
    \begin{eqnarray}\exp(-\beta\Omega W)&=&\int D\Phi D\Phi^\dagger DW_\mu\exp
    \left[\int_\beta dx({\cal L}-J\Phi^\dagger\Phi)\right]\label{minv}\\
    &=&2\pi^2\int\vi^3d\vi\int D\Phi D\Phi^\dagger DW_\mu\,\delta\left(\vi-
    \frac{1}{\beta\Omega}\int_\beta dx\,\vi_1(x)\right)\nonumber\\
    &&\qquad\quad\prod_{i=2}^4\delta\left(\frac{1}{\beta\Omega}\int_\beta dx\,
    \vi_i(x)\right)\exp\left[\int_\beta dx({\cal L}-J\Phi^\dagger\Phi)\right]
    \nonumber\\
    &=&2\pi^2\int\vi^3d\vi\exp\left[-\beta\Omega V(\vi,J)\right]\, .\nonumber
    \end{eqnarray}
Here $\vi_i$ are the real components of $\Phi$ (see eq.~(\ref{higgs}))
and $V(\vi,J)$ is the Landau gauge effective potential of a theory with mass
square $-\nu+J$. In the infinite volume limit only the absolute minimum
$\vi_{min}$ of $V$ contributes to the $\vi$-integral in  eq.~(\ref{minv}),
resulting in the relation
    \be W(J)=V(\vi_{min},J)\, .\label{Vmin}\e
The thermodynamic potential $W(J)$, calculated by this method, is continuous,
but its first derivative has a discontinuity at the critical temperature and
$J=0$, generated by the jump of the absolute minimum $\vi_{min}$ from zero to
the non-trivial, symmetry-breaking minimum $\vi_+$. This discontinuity
corresponds to the shift of the field square expectation value $\sigma$
    \be -\frac{1}{2}\Delta\sigma=\frac{\partial W}{\partial J}\Bigg|_{J=0_-}-
    \frac{\partial W}{\partial J}\Bigg|_{J=0_+}=\frac{\partial}{\partial J}
    \Big[\,V(\vi_+,J)-V(0,J)\,\Big]\, .\e
Considering only the leading terms of the Landau gauge potential $V$, the
approximate result is
    \be -\frac{1}{2}\Delta\sigma=\frac{\partial}{\partial J}\Big[\,\frac{1}{2}
    (-\nu+J+\alpha_{01}T^2)\vi_+^2+\cdots\Big]\approx\frac{1}{2}\vi_+^2\, .\e
This relation justifies the way in which the Landau gauge and the gauge
invariant calculations have been compared in section \ref{gip}, in particular
in figures \ref{potsgi}, \ref{pots} and \ref{phigi}.

The conventional definition of the latent heat from $W(J)$ gives immediately,
via eq.~(\ref{Vmin}), the formula used in the Landau gauge investigation (eq.
\ref{lh})).

Note however, that the gauge invariant $W(J)$ has been calculated in a
completely different manner in section \ref{WJ}. There, it has only played an
intermediate role in the calculation of $V(\sigma)$. The approach was based on
the identification of the minimum of the tree-level potential and a subsequent
loop expansion around this minimum.

In contrast to the paragraph above, a different approach has been taken in this
section. After shifting the scalar field, the effective potential $V(\vi,J)$
which includes loop corrections, is determined. Its minimum then gives the
thermodynamic potential $W(J)$.

The way in which the scalar masses have been resummed in the Landau gauge
calculations of chapters \ref{ahm} and \ref{sm} is not unique. If only the
minima of the potential are to be considered, as suggested by eq.~(\ref{Vmin}),
it may be advantageous to resum differently in the symmetric and in the broken
phase. In particular such a method can take into account the fact that the
Goldstone boson mass has to be zero at the broken minimum. One-loop
calculations, based on this idea, reproduce the numerical results of section
\ref{gip} exactly. A corresponding two-loop investigation is under way
\cite{BFH1}.

The above method, based on the derivatives of $W(J,T)$ near its non-analytic
point, has its disadvantage. No information can be extracted concerning the
metastable and unstable states, thus disabling the usual derivative expansion
approach to the surface tension.

\section{Clausius-Clapeyron equation}
In classical thermodynamics there is a well known relation between the latent
heat of a phase transition and the change of the molar volume, the
Clausius-Clapeyron equation \cite{Cal}. In complete analogy a relation between
latent heat and jump of the order parameter $\sigma$ can be written down.

Describing the state of the system by the gauge invariantly coupled source $J$
and the temperature the phase transition curve is given by a function
$J_{crit.}(T)$. Since $W(J,T)$ is continuous in the $J$-$T$-plane, the total
derivative of $W$ along the phase curve
    \be \frac{dW}{dT}=\frac{\partial W}{\partial J}\,\frac{d J_{crit.}}
    {dT}+\frac{\partial W}{\partial T}\label{WT}\e
is equal in both phases. The partial derivatives of $W$ can be easily related
to the field square expectation value $\sigma$ and the energy density
$E$:
    \be \frac{\partial W}{\partial J}=\frac{1}{2}\sigma\qquad,\qquad\frac{
    \partial W}{\partial T}=\frac{1}{T}(W-E)\, .\e
Now, evaluating eq.~(\ref{WT}) in both phases and equating the results gives
    \be \frac{1}{2}\Delta\sigma\frac{d J_{crit.}}{dT}=\frac{\Delta Q}{T}\, ,\e
where $\Delta Q=E_{symmetric}-E_{broken}$ is the latent heat.

Neglecting effects of higher loops und renormalization $dJ_{crit.}/dT$ can be
easily evaluated by dimensional arguments. Since the Higgs mass term $-\nu+J$
is the only dimensionful parameter of the SU(2)-Higgs model the dependence of
the critical temperature on $J$ is given by
    \be T=\mbox{const.}\,\sqrt{\nu-J_{crit.}}\, .\e
{}From this
    \be T\frac{dJ_{crit.}}{dT}\Bigg|_{J=0}=-2\nu\, \e
and therefore the jump of the order parameter and the latent heat are related
by
    \be \Delta Q=-\frac{1}{2}m_{\mbox{\scriptsize Higgs}}^2\Delta \sigma\,
    ,\label{sq}\e
where $m_{\mbox{\scriptsize Higgs}}$ is the zero temperature Higgs mass.

This relation has been checked against the available data from one- and
two-loop calculations of the effective potential. The Landau gauge results
are in very good agreement with eq.~(\ref{sq}) at small Higgs masses, with
deviations increasing up to $\sim25\%$ at $m_{\mbox{\scriptsize Higgs}}$=80
GeV. It is interesting to observe that the gauge invariant one-loop calculation
of section \ref{gip} satisfies the above relation exactly. This can be proven
analytically, by writing the gauge invariant potential in the form
    \be V(\sigma)=-\frac{\nu}{2}\sigma+T^4\tilde{V}(\sigma/T^2)\e
with some dimensionless function $\tilde{V}$, which has no implicit temperature
dependence. From this general form of $V$ and the definitions of $\Delta\sigma$
and $\Delta Q$ the relation (\ref{sq}) can be easily extracted.

Note also, that the lattice results of ref.~\cite{L2} verify eq.~(\ref{sq})
quite well. For $L_t$=3 lattices the deviation amounts to no more than 1...2
standard deviations, defined by the statistical error of the simulation. Here
$L_t$ denotes the lattice size in time like direction. An improvement of the
validity of eq.~(\ref{sq}) is observed when going from $L_t=2$ to $L_t=3$
lattices.

\chapter*{Conclusions}
\addcontentsline{toc}{chapter}{Conclusions}
The thermodynamic parameters of the electroweak phase transition have been
analysed in a perturbative approach based on the high temperature effective
potential. A complete calculation of the $g^4,\la^2$-potential has been
performed for the Abelian Higgs model, the SU(2)-Higgs model and the standard
model. The Abelian calculation has been shown to have no infrared problems in
the systematic coupling constant expansion in the symmetric phase. This does
not hold in the general case, where the typical non-Abelian diagrams become
important in higher orders. This is seen explicitly at the order $g^4$, where,
opposite to the Abelian case, logarithmic mass terms increase the strength of
the first order phase transition dramatically.

Critical temperature, latent heat, surface tension and jump of the order
parameter $\Phi^\dagger\Phi$ have been calculated for different Higgs masses
and different approximations to the potential. The reliability of the
perturbative expansion is clearly worsening with increasing Higgs mass. In
particular, in the non-Abelian case, any information on an infrared cutoff,
e.g. a magnetic mass value, would increase the accuracy of the calculation
drastically. However, infrared problems are also connected with scalar masses,
which become small at the critical temperature.

Comparing the complete standard model with the pure SU(2)-Higgs model no
qualitative change is found. The main difference is a decrease of the strength
of the phase transition due to the large top quark mass.

Newly available lattice data at $m_{\mbox{\scriptsize Higgs}}\approx 18$~GeV
and at $m_{\mbox{\scriptsize Higgs}}\approx 49$~GeV \cite{L1,L2} are in good
quantitative agreement with the perturbative two-loop results. This may be seen
as a justification to take the calculated parameters more seriously at larger
Higgs mass values, where convergence of the perturbation series is bad and no
lattice data is available. Perturbative results suggest a weak first order
phase transition at realistic Higgs mass values. However, a change to an
analytic crossover is also a possibility \cite{BP}. Unfortunately, the region
where reliable predictions are available and where $\vi\sim T$, an important
condition for baryogenesis, is at $m_{\mbox{\scriptsize Higgs}}\approx 40$~GeV
and therefore well below the experimental bound. The present results seem to
disfavour scenarios of electroweak baryogenesis relying on a strong first order
phase transition within the minimal standard model.

The gauge invariant description, elaborated in the last chapter, allows a
better physical understanding of the thermodynamics of the phase transition. In
this approach, coupling the source term in a gauge invariant manner, a more
direct access to physical quantities is possible. In particular, the symmetric
phase is described by a non-trivial minimum of the potential as well. Since the
numerical results obtained in the gauge invariant approach at one-loop are
similar to the Landau gauge results, the latter ones are strongly supported by
the new, conceptually more satisfactory treatment. Applying the
Clausius-Clapeyron equation to the electroweak phase transition a simple
relation between latent heat and jump of the order parameter has been derived
in the above context. Being in good agreement with perturbative as well as with
lattice data, it improves confidence in the correctness of the treatment of the
phase transition.{\parskip3ex

In} agreement with recent results of other methods the performed investigation
predicts a first order electroweak phase transition of decreasing strength
when the Higgs mass is increasing up to $m_{\mbox{\scriptsize Higgs}}\approx$
70 GeV. At larger Higgs mass values the calculation is strongly affected by
infrared problems. If, in spite of this difficulty, perturbation theory without
explicit infrared cutoff is taken seriously, the phase transition is found to
remain weakly first order.

\newpage
\vspace*{1cm}
\begin{center}{\Large\bf Acknowledgements}
\end{center}
I am most grateful to my advisor, Prof.~W.~Buchm\"uller, who suggested this
investigation, for continuous support and encouragement. During the intensive
and fruitful collaboration, I had many invaluable opportunities to learn from
him. I would like to thank Dr.~Z.~Fodor, with whom I collaborated during large
parts of this investigation, for sharing his ideas with me, for many highly
instructive discussions and, in particular, for patiently explaining the
lattice results to me. At different stages of the work presented in this
thesis, I profited from discussions with Dr.~D.~B\"odeker, J.~Hein,
Dr.~T.~Helbig, Dr.~B.~Kniehl, Dr.~H.-G.~Kohrs and Dr.~O.~Philipsen. I thank
them for their interest in my work. Finally, I would like to thank all my
colleagues and friends at DESY for the enjoyable working atmosphere.

\begin{appendix}
\addcontentsline{toc}{chapter}{Appendix}
\chapter{Some integrals}\label{int}

\section{Integrals of thermal field theory}\label{temp-int}
In this section the basic temperature integrals \cite{AE,P,DJ,AZ}, needed
throughout the calculation, are listed. The notation follows the general
formalism introduced in section \ref{def}. At one-loop level, the fundamental
integral is \cite{AE,DJ}
    \begin{eqnarray}I(m)&=&\is\frac{dk}{k^2+m^2}=\frac{\mu^{2\epsilon}}{(2
    \pi)^{n-1}\beta}\sum_{k_0}\int d^{n-1}k\frac{1}{k^2+m^2}\nonumber\\ \\&=&
    \frac{1}{12\beta^2}(1+\epsilon\ie)-\frac{m}{4\pi\beta}-\frac{m^2}{16\pi^2}
    \left(\frac{1}{\epsilon}+\ln\bar{\mu}^2\beta^2+\frac{3}{2}-c_1\right)+\dots
    \, ,\nonumber\end{eqnarray}
where
    \be c_1=\mif{3}{2}+2\ln{4\pi}-2\gamma_E\approx 5.4076\, .\e
Here terms of higher order in $m\beta$ have been neglected because in the
present calculation they do not contribute to the $g^4$-potential. The
renormalization scale of the $\overline{\mbox{MS}}$-formalism is defined by
    \be \ln\bar{\mu}^2=\ln\mu^2+\ln 4\pi-\gamma_E\, ,\e
where $\gamma_E$ is Euler's constant. The coefficient
    \be \ie=2\gamma_E+\ln\frac{\bar{\mu}^2\beta^2}{4}-\frac{12}{\pi^2}\zeta'(2)
    \label{ie}\e
of $\epsilon$ from the leading contribution
    \be I_\beta^\epsilon=\frac{1}{12\beta^2}(1+\epsilon\ie)\e
will cancel in the final formula for the potential, similarly to the
temperature dependent divergences \cite{AE}.

The fermionic analogue of this integral is
    \begin{eqnarray}I_f(m)&=&\isf\,\,\frac{dk}{k^2+m^2}\nonumber\\ \\&=&-\frac
    {1}{24\beta^2}\left(1+\epsilon(\ie-\ln 4)\right)-\frac{m^2}{16\pi^2}\left(
    \frac{1}{\epsilon}+\ln\bar{\mu}^2\beta^2+\frac{3}{2}-c_1+2\ln 4\right)\, .
    \nonumber\end{eqnarray}
Its leading, mass independent, part will be denoted by $I_{f\beta}^\epsilon$.

Another bosonic one-loop integral, appearing due to the non-covariant structure
of the vector propagator, is given by
    \be \II(m)=\is\frac{k_0^2\,dk}{k^2(k^2+m^2)}=\left(-\frac{1}{2}+\epsilon
    \right)I_\beta^\epsilon-\frac{m^2}{64\pi^2}\left(\frac{1}{\epsilon}+\ln
    \bar{\mu}^2\beta^2+\frac{7}{2}-c_1\right)\, .\e

At two-loop level the basic problem is the calculation of the scalar
setting sun diagram (see \fig\ref{s2l}.a). This has first been done in
ref.~\cite{P} and extended to the case of three different masses in
ref.~\cite{AE}. Neglecting terms of order $m^2\beta^2$ the integral reads
    \begin{eqnarray}H(m_1,m_2,m_3)&=&\is\is\frac{dk\,dq}{(k^2+m_1^2)(q^2+m_2^2)
    ((k+q)^2+m_3^2)}\nonumber\\ \\&=&\frac{3}{16\pi^2\epsilon}I_\beta^\epsilon
    +\frac{1}{64\pi^2\beta^2}\left(\ln\bar{\mu}^2\beta^2-4\ln\frac{\beta(m_1+
    m_2+m_3)}{3}-c_2\right)\, ,\nonumber\end{eqnarray}
where the constant
    \be c_2=2\left(\gamma_E-\frac{\zeta'(2)}{\zeta(2)}+\ln\frac{9}{2}-1\right)
    \approx 3.3025\e
has first been obtained analytically in ref.~\cite{AZ}.

In the standard model calculation the fermionic analogue of this two-loop
integral, i.e. the corresponding integral where two of the propagators have
fermionic Matsubara frequencies $k_0=(2n+1)\pi T$, is needed. However, as has
been shown in ref.~\cite{AE}, this integral vanishes in leading order in
dimensional regularization when $\epsilon\to 0$:
    \be \isf\hspace{.06cm}\is\frac{dk\,dq}{(k^2+m_1^2)(q^2+m_2^2)
    ((k+q)^2+m_3^2)}=\frac{1}{\beta^2}O(m^2\beta^2).\e

\section{One- and two-point functions}\label{AB}
For the convenience of the reader the usual one- and two-point functions,
necessary for the renormalization at $T=0$, are displayed. Working in
$n=4-2\epsilon$ dimensions only the pole and the finite parts are shown.
    \be A(m^2)=\frac{\mu^{2\epsilon}}{(2\pi)^n}\int\frac{d^nk}{k^2-m^2
     +i\varepsilon}=\frac{im^2}{16\pi^2}\left(\frac{1}{\epsilon}+1+\ln
    \frac{\bar{\mu}^2}{m^2}\right)\, ,\e
    \begin{eqnarray} B(q^2,m_0^2,m_1^2)&=&\frac{\mu^{2\epsilon}}{(2\pi)^n}\int
    \frac{d^nk}{(k^2-m_0^2+i\varepsilon)((k+q)^2-m_1^2+i\varepsilon)}\\
    \nonumber\\&
    =&\frac{i}{16\pi^2}\left[\frac{1}{\epsilon}+2+\ln\frac{\bar{\mu}^2}{m_0m_1}
    -\frac{1}{q^2}\left\{(m_1^2-m_0^2)\ln\frac{m_1}{m_0}-2m_1m_0f(x)\right\}
    \right]\, ,\nonumber\end{eqnarray}
where
    \be x=\frac{m_0^2+m_1^2-q^2}{2m_0m_1}\e
and the function $f$ is defined by
    \be f(x)=\left\{\begin{array}{rl@{\quad:\quad}rclrl}&\sqrt{x^2-1}\,
    (-\mbox{arccosh}(-x)+i\pi)&&x&<&-1&\\ \\-&\sqrt{1-x^2}\,\arccos x&-1<&x&<&
    1&\\ \\&\sqrt{x^2-1}\,\,\mbox{arccosh}\,x&1<&x&&&.\end{array}\right.\e

\chapter{Explicit formulae for the effective potential}\label{results}

\section{Abelian Higgs model}\label{ahm-res}
The leading order resummed masses of Goldstone boson, Higgs particle and
longitudinal vector boson read
    \be\begin{array}{rclcl}
    m_\chi^2&=&\la\vi^2-\nu+(4\la+(3-2\epsilon)e^2)I_\beta^\epsilon&=&\la\vi^2-
    \nu+\dfrac{4\la+3e^2}{12\beta^2}+O(\epsilon)\,\, ,\\ \\
    m_\vi^2&=&m_\chi^2+2\la\vi^2\,\, ,\\ \\
    m_L^2&=&e^2\vi^2+4e^2(1-\epsilon)I_\beta^\epsilon&=&e^2\vi^2+\dfrac{e^2}{3
    \beta^2}+O(\epsilon)\,\, .\end{array}\e
The transverse vector mass $m_T=m=e\vi$ remains uncorrected.
For the calculation of $V_R$ the self-energy parts of the type $\Pi_{b2}$ (see
eq.~(\ref{ds2})) are needed. They appear only for the longitudinal and
transverse vector boson:
    \be \Pi_{b2,L}=(2-4\epsilon)e^2I_\beta^\epsilon\quad,\quad
    \Pi_{b2,T}=-2e^2I_\beta^\epsilon\, .\e
The counterterms rendering the potential finite are generated from the leading
order contribution (see the first two terms of eq.~(\ref{V3-ahm}) below):
    \begin{eqnarray}\delta V&=&\frac{\vi^2}{2}\Bigg[-\nu(\delta Z_\nu+\delta
    Z_{\vi^2})+I_\beta^\epsilon\left\{4\la(\delta Z_\la+\delta Z_{\vi^2})
    +(3-2\epsilon)e^2(\delta Z_{e^2}+\delta Z_{\vi^2})\right\}\Bigg]\nonumber\\
    \nonumber\\&&+\frac{\la}{4}\vi^4(\delta Z_\la+2\delta Z_{\vi^2})\, ,
    \label{deltaV}\end{eqnarray}
where
    \be\delta Z_\la=\left(\frac{3e^4}{\la}-6e^2+10\la\right)p\, ,\quad
    \delta Z_{e^2}=\frac{e^2}{3}p\, ,\quad\delta Z_\nu=(4\la-3e^2)p\, ,
    \quad\delta Z_{\vi^2}=3e^2p\, ,\e
with the pole in $\epsilon$ denoted by
    \be p=\frac{1}{16\pi^2\epsilon}\, .\e
The different contributions to the potential, listed below, have to be summed
according to eq.~(\ref{final}), where
    \be V_\ominus=V_a+V_b+V_p\, \e
is given by the diagrams in \fig\ref{u12l}. Linear mass terms, poles in
$\epsilon$ and terms proportional to $\ie$ (see eq.~(\ref{ie})), which cancel
systematically in the final result, are not shown and the limit $\epsilon\to
0$ has already been performed. A finite contribution from $\delta V$ has been
added to $V_4$. The following terms form the complete analytic
$\overline{\mbox{MS}}$-potential:

\begin{eqnarray}
  V_3&\!\!\!=&\!\!\!\frac{\vi^2}{2}\left[-\nu+\frac{1}{\beta^2}\left(\frac{1}
  {3}\la+\frac{1}{4}e^2\right)\right]+\frac{\la}{4}\vi^4-\frac{1}{12\pi\beta}
  \left[m_\vi^3+m_\chi^3+2m^3+m_L^3\right]\label{V3-ahm}
\\
&&\nonumber\\
&&\nonumber\\
  V_a&\!\!\!=&\!\!\!\frac{e^2}{32\pi^2\beta^2}\Bigg[m^2\left(\ln\frac{\beta}{3}
  -\frac{1}{12}\ln\bar{\mu}^2\beta^2-\frac{1}{6}c_1+\frac{1}{4}c_2+\frac{1}{4}
  \right)-m(m_\vi+m_\chi)\\
&&\nonumber\\
  &&\!\!\!\!\!\!\!\!\!\!\!\!+m_\vi m_\chi+\frac{1}{2}(m_\vi^2+m_\chi^2)(-4\ln
  \frac{\beta}{3}+\ln\bar{\mu}^2\beta^2-c_2)-\frac{1}{m^2}(m_\vi^2-m_\chi^2)^2
  \ln(m_\vi+m_\chi)\nonumber\\
&&\nonumber\\
  &&\!\!\!\!\!\!\!\!\!\!\!\!-\frac{1}{m}(m_\vi\!-\!m_\chi)^2(m_\vi+m_\chi)+
  \frac{1}{m^2}\left\{m^4\!-\!2(m_\vi^2+m_\chi^2)m^2+(m_\vi^2\!-\!m_\chi^2)^2
  \right\}\ln(m_\vi+m_\chi+m)\Bigg]\nonumber
\\
&&\nonumber\\
&&\nonumber\\
  V_b&\!\!\!=&\!\!\!\frac{e^2}{64\pi^2\beta^2}\Bigg[5m^2\left(\frac{1}{2}c_2-
  \ln\frac{9\bar{\mu}}{\beta}\right)+\left(8m^2-4m_\vi^2+\frac{m_\vi^4}{m^2}
  \right)\ln(2m+m_\vi)\\
&&\nonumber\\
  &&\!\!\!\!\!\!\!\!\!\!\!\!-\frac{2}{m^2}(m^2-m_\vi^2)^2\ln(m+m_\vi)+4m^2
  \ln(2m_L+m_\vi)+\frac{m_\vi^4}{m^2}\ln m_\vi+2mm_\vi+m_\vi^2\Bigg]\nonumber
\\
&&\nonumber\\
&&\nonumber\\
  V_p&\!\!\!=&\!\!\!-\frac{\la^2\vi^2}{16\pi^2\beta^2}\left[\ln\frac{3\mu^2}
  {\beta^2}-c_2-\ln\{m_\vi^3(m_\vi+2m_\chi)\}\right]
\\
&&\nonumber\\
&&\nonumber\\
  V_4&\!\!\!=&\!\!\!\frac{e^2\vi^2}{64\pi^2\beta^2}\left[\frac{4}{3}\la-
  e^2\left(\frac{35}{18}+\ln\bar{\mu}^2\beta^2-c_1\right)\right]\\
&&\nonumber\\
  &&\!\!\!\!\!\!\!\!\!\!\!\!-\frac{1}{64\pi^2}\left\{\left(m_\vi^4+m_\chi^4
  +3m^4\right)\left(\ln\bar{\mu}^2\beta^2+\frac{3}{2}-c_1\right)-2m^4
  \right\}\nonumber
\\
&&\nonumber\\
&&\nonumber\\
  V_z&\!\!\!=&\!\!\!\frac{1}{32\pi^2\beta^2}\left[\frac{3}{2}\la(m_\vi^2
  +m_\chi^2)+\la m_\vi m_\chi+e^2(m_L+2m)(m_\vi+m_\chi)\right]
\end{eqnarray}\\[-1cm]

\section{Standard model}\label{sm-res}
The leading order resummed masses of Goldstone boson and Higgs particle read
    \be\begin{array}{rclcl}
    m_2^2&=&\la\vi^2-\nu+\left(6\la+\dfrac{3g_2^2+g_1^2}{4}(3-2\epsilon)\right)
    I_\beta^\epsilon-6g_Y^2I_{f\beta}^\epsilon &&\\ \\&=&\la\vi^2-\nu+\dfrac{1}
    {12\beta^2}\left(6\la+\dfrac{9}{4}g_2^2+\dfrac{3}{4}g_1^2+3g_Y^2
    \right)+O(\epsilon)\,\, ,&&\\ \\
    m_1^2&=&m_2^2+2\la\vi^2\,\, .\end{array}\label{msm}\e
While the transverse vector boson masses and the fermion mass remain
uncorrected to leading order
    \be m_W=\frac{1}{2}g_2\vi\quad,\quad m_Z=m_W/\cos \theta_W\quad,
    \quad m_f=\frac{1}{\sqrt{2}}g_Y\vi\, ,\e
the longitudinal SU(2)$\times$U(1) mass matrix receives temperature
corrections in the diagonal elements \cite{Ca}
    \be\begin{array}{rclcl}\delta m_{WL}^2&=&g_2^2\left[(10-18\epsilon)
    I_\beta^\epsilon-8n_f(1-\epsilon)I_{f\beta}^\epsilon\right]&=&\dfrac{g_2^2}
    {\beta^2}\left(\dfrac{5}{6}+\dfrac{1}{3}n_f\right)+O(\epsilon)\,\, ,\\ \\
    \delta m_{BL}^2&=&g_1^2\left[(2-2\epsilon)I_\beta^\epsilon-\dfrac{40}{3}n_f
    (1-\epsilon)I_{f\beta}^\epsilon\right]&=&\dfrac{g_1^2}{\beta^2}\left(
    \dfrac{1}{6}+\dfrac{5}{9}n_f\right)+O(\epsilon)\,\, .\end{array}\e
They result in longitudinal masses defined by\\[.1cm]
    \be \!\!\left(\!\begin{array}{cc}
        m_W^2+\delta m_{WL}^2 & -\mif{1}{4}g_1g_2\vi^2  \\&\\
       -\mif{1}{4}g_1g_2\vi^2          & m_B^2+\delta m_{BL}^2
    \end{array}\!\right)\!\!       =\!\!
    \left(\!\begin{array}{cc}
        \cos\tilde{\theta} & \sin\tilde{\theta} \\&\\
       -\sin\tilde{\theta} & \cos\tilde{\theta} \end{array}\!\right)\!\!
    \left(\!\begin{array}{cc}
        m_{ZL}^2 & 0 \\&\\ 0 & m_{\gamma L}^2 \end{array}\!\right)\!\!
    \left(\!\begin{array}{cc}
        \cos\tilde{\theta} & -\sin\tilde{\theta} \\&\\
        \sin\tilde{\theta} & \cos\tilde{\theta}\end{array}\!\right)\!.
    \label{mm}\e\\[.1cm]
In the following the short hand notations
    \be s=\sin \theta_W\quad,\quad c=\cos \theta_W\quad,\quad \tilde{s}=\sin
    \tilde{\theta}\quad,\quad \tilde{c}=\cos \tilde{\theta} \e
are used. From eq.~(\ref{mm}) several helpful identities can be derived:
    \begin{eqnarray}m_{ZL}^2m_{\gamma L}^2&=&m_Z^2\left[c^2\delta m_{BL}^2+s^2
    \delta m_{WL}^2\right]+\delta m_{BL}^2\delta m_{WL}^2\nonumber\\
    m_{ZL}^2+m_{\gamma L}^2&=&m_Z^2+\delta m_{WL}^2+\delta m_{BL}^2\label{^4}\\
    m_{ZL}^4+m_{\gamma L}^4&=&\left[m_Z^2+\delta m_{WL}^2+\delta m_{BL}^2
    \right]^2-2m_Z^2\left[c^2\delta m_{BL}^2+s^2\delta m_{WL}^2\right]\, .
    \nonumber\end{eqnarray}
For the calculation of $V_R$ self-energy parts of the type $\Pi_{a2}(k)$ and
$\Pi_{b2}$ are needed:
    \be\begin{array}{lllllll}
    \Pi_{b2}^{scalars}&=&-6g_Y^2I_{f\beta}^\epsilon&,&
    \Pi_{a2,L}^{WW}(k)&=&2g^2I_\beta^\epsilon(1-\epsilon)\dfrac{k_0^2}{k^2}\, ,
    \\ \\ \Pi_{b2,T}^{WW}&=&-g_2^2(6-4\epsilon)I^\epsilon_\beta&,&
    \Pi_{b2,L}^{WW}&=&g_2^2\left[2(3-8\epsilon)I_\epsilon^\beta-8n_f(1-\epsilon
    )I_{f\beta}^\epsilon\right]\\ \\
    \Pi_{b2,T}^{BB}&=&-g_1^2I^\epsilon_\beta&,&\Pi_{b2,L}^{BB}&=&g_1^2\left[
    (1-2\epsilon)I_\epsilon^\beta-\dfrac{40}{3}n_f(1-\epsilon)
    I_{f\beta}^\epsilon\right]\end{array}\, .\e
In complete analogy with the Abelian case (see eq.~(\ref{deltaV})) the
counterterms are generated from the leading order potential given by the first
line of eq.~(\ref{V3-sm}). The multiplicative renormalization realizing the
$\overline{\mbox{MS}}$-scheme is described by

    \[\delta Z_{\vi^2}=\left(\frac{9}{4}g_2^2+\frac{3}{4}g_1^2-3g_Y^2
    \right)p\, ,\quad
    \delta Z_{g_1^2}=g_1^2\left(\frac{1}{6}+\frac{20}{9}n_f\right)p\, ,\quad
    \delta Z_{g_2^2}=g_2^2\left(-\frac{43}{6}+\frac{4}{3}n_f\right)p\, ,\]
    \[\]
    \be \delta Z_\la=\frac{1}{\la}\left(-3g_Y^4+\frac{3}{16}g_1^4+\frac{3}{8}
    g_1^2g_2^2+\frac{9}{16}g_2^4+6\la g_Y^2-\frac{3}{2}\la g_1^2-\frac{9}{2}\la
    g_2^2+12\la^2\right)p\, ,\label{ct-sm}\e\[\]
    \[\delta Z_{g_Y}=\left(\frac{9}{4}g_Y^2-\frac{17}{24}g_1^2-
    \frac{9}{8}g_2^2-4g_s^2\right)p\qquad ,\qquad
    \delta Z_\nu=\left(3g_Y^2-\frac{3}{4}g_1^2-\frac{9}{4}g_2^2+6\la\right)
    p\, .\]
As in the appendix \ref{ahm-res}
the final result is presented without
linear mass terms, poles and $\ie$-terms, which would cancel in the sum. The
following contributions form the complete standard model effective potential
to the order $g^4,\la^2$:

\begin{eqnarray}
  V_3&\!\!\!=&\!\!\!\frac{\vi^2}{2}\left[-\nu+\frac{1}{\beta^2}\left(\frac{1}
  {2}\la+\frac{3}{16}g_2^2+\frac{1}{16}g_1^2+\frac{1}{4}g_Y^2\right)\right]
  +\frac{\la}{4}\vi^4\label{V3-sm}\\
&&\nonumber\\
  &&\!\!\!\!\!\!\!\!\!\!\!\!-\frac{1}{12\pi\beta}\left[{ m_1}^{3}+3\,
  { m_2}^{3}+4\,{ m_W}^{3}+2\,{ m_{WL}}^{3}+2\,{ m_Z}^{3}+{ m_{ZL}}^{3}+
  { m_{\gamma L}}^{3}\right]\nonumber
\\
&&\nonumber\\
&&\nonumber\\
  V_a&\!\!\!=&\!\!\!\frac{g_2^2}{32\pi^2\beta^2}\;\Bigg[m_W^2\left(2-\frac{1}
  {c^2}+\frac{1}{2c^4}\right)\left(\ln{\frac{\beta}{3}}-\frac{1}{12}\ln{
  \bar{\mu}^2\beta^2}-\frac{1}{6}c_1+\frac{1}{4}c_2+\frac{1}{4}\right)\\
&&\nonumber\\
  &&\!\!\!\!\!\!\!\!\!\!\!\!+\frac{1}{4}\left(1+\frac{1}{2c^2}\right)\left\{
  \left(m_1^2+3m_2^2\right)\left(-4\ln{\frac{\beta}{3}}+\ln{\bar{\mu}^2\beta^2}
  -c_2\right)+2m_2\left(m_1+m_2\right)\right\}\nonumber\\
&&\nonumber\\
  &&\!\!\!\!\!\!\!\!\!\!\!\!-4s^2m_2^2\ln{(2m_2)}-\frac{1}{2m_W}\left(1+\frac
  {1}{2c}\right)(m_1-m_2)^2(m_1+m_2)+2m_2m_Zs^2\nonumber\\
&&\nonumber\\
  &&\!\!\!\!\!\!\!\!\!\!\!\!-\frac{m_W}{2}\left(1+\frac{1}{2c^3}\right)
  (m_1+3m_2)-\frac{3}{4m_W^2}\left(m_1^2-m_2^2\right)^2\ln(m_1+m_2)\nonumber\\
&&\nonumber\\
  &&\!\!\!\!\!\!\!\!\!\!\!\!+\frac{1}{2m_W^2}\left\{m_W^4-2\left(m_1^2+m_2^2
  \right)m_W^2+\left(m_1^2-m_2^2\right)^2\right\}\ln{(m_1+m_2+m_W)}\nonumber\\
&&\nonumber\\
  &&\!\!\!\!\!\!\!\!\!\!\!\!+\frac{1}{4m_W^2}\left\{m_Z^4-2\left(m_1^2+m_2^2
  \right)m_Z^2+\left(m_1^2-m_2^2\right)^2\right\}\ln{(m_1+m_2+m_Z)}\nonumber\\
&&\nonumber\\
  &&\!\!\!\!\!\!\!\!\!\!\!\!+\left(\frac {1}{4c^2}-s^2\right)\left(m_Z^2-4
  m_2^2\right)\ln(2m_2+m_Z)+\frac{1}{2}\left(m_W^2-4m_2^2\right)\ln(2m_2+m_W)
  \Bigg]\nonumber
\\
&&\nonumber\\
&&\nonumber\\
V_b&\!\!\!=&\!\!\!\frac{g_2^2}{64\pi^2\beta^2}\Bigg[\left\{\left(c^4+\frac{1}
  {c^4}+4c^2+\frac{4}{c^2}-10\right)m_W^2+2\left(c^2-\frac{1}{c^2}\right)s^2
  m_2^2+\frac{s^4m_2^4}{m_W^2}\right\}\\
&&\nonumber\\
  &&\!\!\!\!\!\!\!\!\!\!\!\!\times\ln(m_W+m_Z+m_2)+\left\{\left(5-4c^2\right)
  m_W^2-\frac{1}{m_W^2}\left(m_W^2c^2+m_2^2s^2\right)^2\right\}\ln(m_W+m_2)
  \nonumber\\
&&\nonumber\\
  &&\!\!\!\!\!\!\!\!\!\!\!\!-\frac{s^4}{m_W^2}\left(m_Z^2-m_2^2\right)^2\ln(
  m_Z+m_2)+m_W\left\{m_2\left(\frac{1}{c^2}-c^2+\frac{s^4}{c}
  \right)+m_1\left(1+\frac{1}{2c^3}\right)\right\}\nonumber\\
&&\nonumber\\
  &&\!\!\!\!\!\!\!\!\!\!\!\!+m_W^2\left\{\left(\frac {5}{2c^2}+\frac{5}{8c^4}
  -\frac{5}{4}\right)\left(2\ln\frac{\beta}{9\bar{\mu}}+c_2\right)+
  \frac{c^2}{2}-\frac{5}{2}+\frac{2}{c^2}+\frac{s^2}{c}\left(c^2-
  \frac{1}{c^2}\right)\right\}\nonumber\\
&&\nonumber\\
  &&\!\!\!\!\!\!\!\!\!\!\!\!-\frac{1}{2m_W^2}\left\{2\left(m_W^2-m_1^2\right)^2
  \ln(m_W+m_1)+\left(m_Z^2-m_1^2\right)^2\ln(m_Z+m_1)\right\}\nonumber\\
&&\nonumber\\
  &&\!\!\!\!\!\!\!\!\!\!\!\!+\!\!\left(4m_W^2-2m_1^2+\frac
  {m_1^4}{2m_W^2}\right)\ln(2m_W+m_1)+\frac{1}{c^2}\left(2m_Z^2
  -m_1^2+\frac{m_1^4}{4m_Z^2}\right)\ln(2m_Z+m_1)\nonumber\\
&&\nonumber\\
  &&\!\!\!\!\!\!\!\!\!\!\!\!+s^2m_2^2\left(2\ln m_2+
  \frac{s^2}{c}\right)+\frac{m_1^2}{2}\left(1+\frac{1}{2c^2}\right)+\frac{1}
  {m_W^2}\left(s^4m_2^4\ln m_2+\frac{3}{4}m_1^4\ln m_1\right)\nonumber\\
&&\nonumber\\
  &&\!\!\!\!\!\!\!\!\!\!\!\!
  +\frac{\vi^2}{4g_2^2}\Big\{(g_2\tilde{c}+g_1\tilde{s})^4\ln(2m_{ZL}+m_1)
  +(g_2\tilde{s}-g_1\tilde{c})^4\ln(2m_{\gamma L}+m_1)\nonumber\\
&&\nonumber\\
  &&\!\!\!\!\!\!\!\!\!\!\!\!+2g_2^4\ln(2m_{WL}+m_1)
  +2(g_2\tilde{c}+g_1\tilde{s})^2(g_2\tilde{s}-g_1\tilde{c})^2\ln(m_{ZL}+m_{
  \gamma L}+m_1)\nonumber\\
&&\nonumber\\
  &&\!\!\!\!\!\!\!\!\!\!\!\!
  +4g_2^2g_1^2\left(\tilde{s}^2\ln(m_{WL}+m_{ZL}+m_2)
  +\tilde{c}^2\ln(m_{WL}+m_{\gamma L}+m_2)\right)\Big\}\Bigg]\nonumber
\\
&&\nonumber\\
&&\nonumber\\
  V_i&\!\!\!=&\!\!\!\frac{1}{8\pi^2\beta^2}\Bigg[\left\{\frac{g_2^2m_f^2}{96}
  \left(10+\frac{17}{c^2}\right)+\frac{g_2^2n_fm_W^2}{36}\left(\frac{10}{c^2}
  -\frac{5}{c^4}-14\right)+g_s^2m_f^2\right\}\\
&&\nonumber\\
  &&\!\!\!\!\!\!\!\!\!\!\!\!\times(\ln{\bar{\mu}^2\beta^2}-c_1+\frac{1}{2}
  +\frac{10}{3}\ln2)-\frac{4g_2^2n_fm_W^2}{27}\left(\frac{10}{c^2}-\frac{5}
  {c^4}-14\right)\ln2\Bigg]\nonumber
\\
&&\nonumber\\
&&\nonumber\\
  V_j&\!\!\!=&\!\!\!\frac{g_Y^2}{128\pi^2\beta^2}\left[\left(9m_f^2-m_1^2-
  3m_2^2\right)\left(\ln{\bar{\mu}^2\beta^2}-c_1+\frac{3}{2}-\ln4\right)
  +48m_f^2\ln2\right]
\\
&&\nonumber\\
&&\nonumber\\
  V_m&\!\!\!=&\!\!\!\frac{g_2^2}{16\pi^2\beta^2}\Bigg[m_W^2\Bigg(\Big(-\frac
  {1}{4c^4}-\frac{1}{c^2}+\frac{5}{2}-c^2-\frac{c^4}{4}\Big)\ln(m_W+m_Z)\\
&&\nonumber\\
  &&\!\!\!\!\!\!\!\!\!\!\!\!+\left(\frac{1}{8c^4}+\frac{1}{c^2}-5-4c^2\right)
  \ln(2m_W+m_Z)+\frac{31}{8}\ln{\bar{\mu}^2\beta^2}-\frac{11}{16}c_1-\frac
  {51}{16}c_2-\frac{251}{96}\nonumber\\
&&\nonumber\\
  &&\!\!\!\!\!\!\!\!\!\!\!\!-\frac{11}{12}c-\frac{5}{4c}+\frac{1}{8c^2}-4s^2
  \ln2+\frac{1}{4}c^2(c-\frac{1}{2})+\frac{1}{8}\left(2-\frac{1}{c^4}\right)
  \ln c-\frac{51}{4}\ln\frac{\beta}{3}\nonumber\\
&&\nonumber\\
  &&\!\!\!\!\!\!\!\!\!\!\!\!+\left(\frac{1}{8c^4}-\frac{23}{4}+5c^2+\frac{1}
  {4}c^4\right)\ln m_W\Bigg)-m_Wm_{WL}(1+c)-2s^2m_{WL}^2\ln(2m_{WL})
  \nonumber\\
&&\nonumber\\
  &&\!\!\!\!\!\!\!\!\!\!\!\!+\left(\frac{1}{2}m_W^2-2c^2m_{WL}^2\right)\ln(2
  m_{WL}+m_Z)+\frac{1}{2}m_{WL}^2-\frac{m_{WL}^3}{m_W}\nonumber\\
&&\nonumber\\
  &&\!\!\!\!\!\!\!\!\!\!\!\!+\tilde{s}^2\Bigg\{\left(m_W^2-2m_{WL}^2-
  2m_{\gamma L}^2+\frac{\left(m_{WL}^2-m_{\gamma L}^2\right)^2}{m_W^2}\right)
  \ln(m_{WL}+m_{\gamma L}+m_W)\nonumber\\
&&\nonumber\\
  &&\!\!\!\!\!\!\!\!\!\!\!\!+m_{\gamma L}(m_{WL}-m_W)+\frac{
  m_{WL}^2m_{\gamma L}+m_{WL}m_{\gamma L}^2-m_{\gamma L}^3}{m_W}
  -\frac{\left(m_{WL}^2-m_{\gamma L}^2\right)^2\ln(m_{\gamma L}
  +m_{WL})}{m_W^2}\Bigg\}\nonumber\\
&&\nonumber\\
  &&\!\!\!\!\!\!\!\!\!\!\!\!+\tilde{c}^2\Bigg\{\hspace{1cm}
  m_{\gamma L}\longrightarrow m_{ZL} \hspace{1cm}\Bigg\}\Bigg]\nonumber
\\
&&\nonumber\\
&&\nonumber\\
  V_p&\!\!\!=&\!\!\!-\frac{3\la^2\vi^2}{32\pi^2\beta^2}\left[\ln
  \frac{9\bar{\mu}^2}{\beta^2}-c_2-2\ln\{m_1(m_1+2m_2)\}\right]
\\
&&\nonumber\\
&&\nonumber\\
  V_4&\!\!\!=&\!\!\!\frac{\vi^2}{64\pi^2\beta^2}\Bigg[\frac{g_2^4}{4}\left\{
  \left(\frac{1}{c^2}-\frac{1}{2c^4}-\frac{35}{4}\right)(
  \ln{\bar{\mu}^2\beta^2}-c_1)+\frac{293}{72}-\frac{1}{18c^2}
  -\frac{13}{18c^4}\right\}\label{V4sm}\\
&&\nonumber\\
  &&\!\!\!\!\!\!\!\!\!\!\!\!+\frac{g_2^4n_f}{27}\left(\frac{10}{c^2}
  -\frac{5}{c^4}-14\right)+g_Y^2\left(\frac{g_2^2s^2}{3c^2}-\frac{3g_Y^2}{4}
  +4g_s^2\right)\ln4+g_2^2\left(\frac{g_Y^2}{2}+\la\right)\left(1+\frac{1}
  {2c^2}\right)\nonumber\\
&&\nonumber\\
  &&\!\!\!\!\!\!\!\!\!\!\!\!+3g_Y^2\la\left(\ln{\bar{\mu}^2\beta^2}-c_1+\frac
  {3}{2}\right)\Bigg]-\frac{1}{64\pi^2}\Big\{-4m_W^4-2m_Z^4-24m_f^4\ln4
  \nonumber\\
&&\nonumber\\
  &&\!\!\!\!\!\!\!\!\!\!\!\!+\left(6m_W^4
  +3m_Z^4+m_1^4+3m_2^4-12m_f^4\right)\left(
  \ln{\bar{\mu}^2\beta^2}-c_1+\frac{3}{2}\right)\Big\}\nonumber
\\
&&\nonumber\\
&&\nonumber\\
  V_z&\!\!\!=&\!\!\!\frac{1}{32\pi^2\beta^2}\Bigg[\mif{1}{4}(m_1+3m_2)\Big\{
  g_2^2\left(2m_{WL}+4m_W+m_{ZL}\tilde{c}^2+2m_Zc^2\right)\\
&&\nonumber\\
  &&\!\!\!\!\!\!\!\!\!\!\!\!+g_1^2\left(m_{ZL}\tilde{s}^2+2m_Zs^2\right)
  +m_{\gamma L}\left(g_2^2\tilde{s}^2+g_1^2\tilde{c}^2\right)\Big\}\nonumber\\
&&\nonumber\\
  &&\!\!\!\!\!\!\!\!\!\!\!\!+\mif{1}{2}g_2g_1(m_1-m_2)\left\{2m_Zsc+(m_{ZL}
  -m_{\gamma L})\tilde{s}\tilde{c}\right\}+4g_2^2m_{WL}\left(m_W
  +m_Zc^2\right)\nonumber\\
&&\nonumber\\
  &&\!\!\!\!\!\!\!\!\!\!\!\!+g_2^2m_W\left(4m_{\gamma L}\tilde{s}^2+
  \mif{8}{3}m_W+\mif{16}{3}m_Zc^2+4m_{ZL}\tilde{c}^2\right)+3\la\left(
  m_1m_2+\mif{1}{2}m_1^2+\mif{5}{2}m_2^2\right)\Bigg].\nonumber
\end{eqnarray}\\[-1cm]

\newpage
\section{SU(2)-Higgs model}\label{su2-res}
In this section, for completeness, the potential of the SU(2)-Higgs model to
order $g^4,\la^2$ is presented. It is readily obtained from the standard model
results of section \ref{sm-res} by performing the limit $g_1,g_Y\to 0$ and
setting the number of families $n_f$ to zero. With the masses
    \be\begin{array}{lllllll}m_2^2&=&\la\vi^2-\nu+\dfrac{1}{12\beta^2}\left(6
    \la+\dfrac{9}{4}g^2\right)&\quad,\quad&m_1^2&=&m_2^2+2\la\vi^2\, ,\\ \\
    m_L^2&=&\dfrac{1}{4}g^2\vi^2+\dfrac{5}{6}\dfrac{g^2}{\beta^2}&\quad,
    \quad&m^2&=&\dfrac{1}{4}g^2\vi^2\, ,\end{array}\e
the final result is given by the sum of the following contributions:

\begin{eqnarray}
  V_3&\!\!\!=&\!\!\!\frac{\vi^2}{2}\left[-\nu+\frac{1}{\beta^2}\left(\frac{1}
  {2}\la+\frac{3}{16}g^2\right)\right]+\frac{\la}{4}\vi^4-\frac{1}{12\pi\beta}
  \left[{m_1}^{3}+3\,{m_2}^{3}+6\,{ m}^{3}+3\,{ m_{L}}^{3}\right]\label{V3-su2}
  \\
&&\nonumber\\
&&\nonumber\\
  V_a&\!\!\!=&\!\!\!\frac{3g^2}{128\pi^2\beta^2}\;\Bigg[2m^2\left(\ln{\frac{
  \beta}{3}}-\frac{1}{12}\ln{\bar{\mu}^2\beta^2}-\frac{1}{6}c_1+\frac{1}{4}c_2+
  \frac{1}{4}\right)+m_2\left(m_1+m_2\right)\\
&&\nonumber\\
  &&\!\!\!\!\!\!\!\!\!\!\!\!+\frac{1}{2}\left(m_1^2+3m_2^2\right)\left(-4\ln{
  \frac{\beta}{3}}+\ln{\bar{\mu}^2\beta^2}-c_2\right)-\frac{1}{m}(m_1-m_2)^2
  (m_1+m_2)-m(m_1+3m_2)\nonumber\\
&&\nonumber\\
  &&\!\!\!\!\!\!\!\!\!\!\!\!-\frac{1}{m^2}\left(m_1^2-m_2^2\right)^2\ln(m_1+
  m_2)+\left(m^2-4m_2^2\right)\ln(2m_2+m)\nonumber\\
&&\nonumber\\
  &&\!\!\!\!\!\!\!\!\!\!\!\!+\frac{1}{m^2}\left\{m^4-2\left(m_1^2+m_2^2\right)
  m^2+\left(m_1^2-m_2^2\right)^2\right\}\ln{(m_1+m_2+m)}\Bigg]\nonumber
\\
&&\nonumber\\
&&\nonumber\\
V_b&\!\!\!=&\!\!\!\frac{3g^2}{64\pi^2\beta^2}\Bigg[m^2\left(\frac{5}{4}\ln
  \frac{\beta}{9\bar{\mu}}+\frac{5}{8}c_2+\ln(2m_L+m_1)\right)+\frac{m_1^4}
  {4m^2}\ln m_1+\frac{1}{2}m_1m\\
&&\nonumber\\
  &&\!\!\!\!\!\!\!\!\!\!\!\!-\frac{1}{2m^2}(m^2-m_1^2)^2\ln(m+m_1)+\left(2m^2
  -m_1^2+\frac{m_1^4}{4m^2}\right)\ln(2m+m_1)+\frac{1}{4}m_1^2\Bigg]\nonumber
\\
&&\nonumber\\
&&\nonumber\\
  V_m&\!\!\!=&\!\!\!\frac{g^2}{16\pi^2\beta^2}\Bigg[\frac{m^2}{8}\left(31\ln
  \bar{\mu}^2\beta^2-66\ln m+39\ln 3-\frac{11}{2}c_1-\frac{51}{2}c_2-\frac{145}
  {4}-102\ln\beta\right)\nonumber\\
&&\nonumber\\
  &&\!\!\!\!\!\!\!\!\!\!\!\!-3m_Lm+\frac{3}{2}m_L^2+\left(\frac{3}{2}m^2-6m_L^2
  \right)\ln(2m_L+m)\Bigg]
\\
&&\nonumber\\
&&\nonumber\\
  V_p&\!\!\!=&\!\!\!-\frac{3\la^2\vi^2}{32\pi^2\beta^2}\left[\ln
  \frac{9\bar{\mu}^2}{\beta^2}-c_2-2\ln\{m_1(m_1+2m_2)\}\right]
\\
&&\nonumber\\
&&\nonumber\\
  V_4&\!\!\!=&\!\!\!\frac{m^2}{64\pi^2\beta^2}\left[g^2\left\{\frac{79}{24}-
  \frac{33}{4}(\ln\bar{\mu}^2\beta^2-c_1)\right\}+6\la\right]\\
&&\nonumber\\
  &&\!\!\!\!\!\!\!\!\!\!\!\!-\frac{1}{64\pi^2}\left\{\left(9m^4+m_1^4+3m_2^4
  \right)\left(\ln\bar{\mu}^2\beta^2-c_1+\frac{3}{2}\right)-6m^4\right\}
  \nonumber
\\
&&\nonumber\\
&&\nonumber\\
  V_z&\!\!\!=&\!\!\!\frac{1}{32\pi^2\beta^2}\Bigg[\frac{g^2}{4}(m_1+3m_2)(3m_L+
  6m)+g^2m(8m+12m_L)\\
&&\nonumber\\
  &&\!\!\!\!\!\!\!\!\!\!\!\!+3\la(m_1m_2+\frac{1}{2}m_1^2+\frac{5}{2}m_2^2)
  \Bigg]\, .
\end{eqnarray}

\chapter{Formulae for renormalization at $T=0$}\label{se}
\section{Abelian Higgs model}\label{ahm-se}
The self-energy of the vector field (in Minkowski space) has the structure
\be i\Pi^{\mu\nu}(q^2)=i\left(\Pi(q^2)g^{\mu\nu}+\Pi_Lq^\mu q^\nu\right)\, .\e
Here the momentum independent part of $\Pi(q^2)$ is just the correction to the
mass, and $\Pi_L(q^2)$ is irrelevant in Landau gauge because the propagator is
transverse. The complete one-loop result without one-particle reducible
tadpoles reads
    \begin{eqnarray}
    \back\Pi(q^2)&=&ie^2A(m_\vi^2)+\frac{ie^2}{3-2\epsilon}\Bigg\{
    \left(\frac{m^2-m_\vi^2}{q^2}-1\right)A(m_\vi^2)+\left(\frac{m_\vi^2-m^2}
    {q^2}-1\right)A(m^2)\nonumber\\ \\
    \back&&+\left[(1-\epsilon)8m^2+\frac{(m_\vi^2-m^2-q^2)^2}{q^2}\right]
    B(q^2,m_\vi^2,m^2)\Bigg\}\, .\nonumber\end{eqnarray}
Similarly, the one-loop self-energy of the Higgs particle is given by
    \begin{eqnarray}\bac\Pi_\vi(q^2)&=&3i\la A(m_\vi^2)+\left(3-2\epsilon-
    \frac{q^2}{m^2}\right)ie^2A(m^2)+\frac{ie^2}{2m^2}(m_\vi^4-q^4)B(q^2,0,0)
    \nonumber\\ \\
    \bac&&+ie^2\left[(3-2\epsilon)2m^2-2q^2+\frac{q^4}{2m^2}\right]B(q^2,m^2,
    m^2)+18i\la^2v^2B(q^2,m_\vi^2,m_\vi^2)\, .\nonumber\end{eqnarray}
Here $A(q^2)$ and $B(q^2,m_0^2,m_1^2)$ are the usual one- and two-point
functions, defined in appendix \ref{AB}.
\pagebreak

\section{Standard model}\label{sm-se}
The Landau gauge self-energy of the Higgs particle is given by
\begin{eqnarray}
  \Pi_\vi(q^2)&\!\!\!=&\!\!\!3i\la A(M_H^2)+i\la^2v^2\left[18B(q^2,M_H^2,
  M_H^2)+6B(q^2,0,0)\right]\\
&&\nonumber\\
  &&\!\!\!\!\!\!\!\!\!\!\!\!+\frac{ig_2^4v^2}{8}\Bigg[2\Bigg\{\frac{1}
  {2M_W^2}A(M_W^2)+\left(3-2\epsilon-\frac{q^2}{M_W^2}+\frac{q^4}{4M_W^4}
  \right)B(q^2,M_W^2,M_W^2)\nonumber\\
&&\nonumber\\
  &&\!\!\!\!\!\!\!\!\!\!\!\!+\left(-\frac{1}{2}+\frac{q^2}{M_W^2}-\frac{q^4}
  {2M_W^4}\right)B(q^2,M_W^2,0)+\frac{q^4}{4M_W^4}B(q^2,0,0)\Bigg\}+\frac{1}
  {c^4}\Bigg\{\,M_W\to M_Z\,\Bigg\}\Bigg]\nonumber\\
&&\nonumber\\
  &&\!\!\!\!\!\!\!\!\!\!\!\!-\frac{ig_2^2}{4}\Bigg[2\Bigg\{\left(-2+2\epsilon+
  \frac{q^2}{M_W^2}\right)A(M_W^2)-\frac{(q^2-M_W^2)^2}{M_W^2}B(q^2,M_W^2,0)+
  \frac{q^4}{M_W^2}B(q^2,0,0)\Bigg\}\nonumber\\
&&\nonumber\\
  &&\!\!\!\!\!\!\!\!\!\!\!\!+\frac{1}{c^2}\Bigg\{\,M_W\to M_Z\,\Bigg\}\Bigg]-3i
  g_Y^2\left[2A(m_t^2)+(4m_t^2-q^2)B(q^2,m_t^2,m_t^2)\right]\,.\nonumber
\end{eqnarray}
Here $M_W\, ,M_Z$ and $m_t$ are the zero temperature masses of the vector
bosons and the top quark and $c$ is the cosine of the Weinberg angle.

The fermion self energy can be written in the form
    \be \Sigma^f(q)=\Sigma^f_S(q^2)+\Sigma^f_V(q^2)\slq+\Sigma^f_A(q^2)\slq\;
    \gamma_5\, .\e
The $\gamma_5$-contribution is irrelevant for the mass counterterm $\delta m_t
=\Sigma^f_S(m_t^2)+m_t\Sigma^f_V(m_t^2)$, which is obtained from
\begin{eqnarray}
  \Sigma^f_S(m_t^2)&\!\!\!=&\!\!\!-\frac{ig_Y^2m_t}{2}[B(m_t^2,m_t^2,M_H^2)-
  B(m_t^2,m_t^2,0)]-\frac{4}{3}g_s^2I_S(0,m_t^2)\\
&&\nonumber\\
  &&\!\!\!\!\!\!\!\!\!\!\!\!-\frac{g_2^2}{9}\left[4s^2I_S(0,m_t^2)-\frac{s^2}
  {c}\left(3c-\frac{s^2}{c}\right)I_S(M_Z^2,m_t^2)\right]\, ,\nonumber
\\
&&\nonumber\\
&&\nonumber\\
  \Sigma^f_V(m_t^2)&\!\!\!=&\!\!\!\frac{ig_Y^2}{4m_t^2}\Bigg[2A(m_t^2)-A(M_H^2)
  +(M_H^2-2m_t^2)B(m_t^2,m_t^2,M_H^2)-2m_t^2B(m_t^2,m_t^2,0)\nonumber\\
&&\nonumber\\
  &&\!\!\!\!\!\!\!\!\!\!\!\!-m_t^2B(m_t^2,0,0)\Bigg]-\frac{4}{3}g_s^2I_V(0,
  m_t^2)\\
&&\nonumber\\
  &&\!\!\!\!\!\!\!\!\!\!\!\!-g_2^2\Bigg[\frac{1}{4}I_V(M_W^2,0)+\frac{4}{9}s^2
  I_V(0,m_t^2)+\left\{\frac{2s^4}{9c^2}+\frac{1}{72}\left(3c-\frac{s^2}{c}
  \right)^2\right\}I_V(M_Z^2,m_t^2)\Bigg]\, .\nonumber
\end{eqnarray}
Here the following short forms for typical self-energy contributions have been
used:
\begin{eqnarray}
  I_S(m_0^2,m_1^2)&\!\!\!=&\!\!\!-im_1(3-2\epsilon)B(m_t^2,m_0^2,m_1^2)
\\
&&\nonumber\\
  I_V(m_0^2,m_1^2)&\!\!\!=&\!\!\!\frac{i}{2m_t^2m_0^2}\Bigg[\left\{(2-2\epsilon
  )m_0^2+m_1^2-
  m_t^2\right\}A(m_0^2)-(2-2\epsilon)m_0^2A(m_1^2)\\
&&\nonumber\\
  &&\!\!\!\!\!\!\!\!\!\!\!\!+\Bigg\{2m_t^2m_0^2(1-2\epsilon)+(3-2\epsilon)m_0^2
  \left(m_1^2-m_0^2-m_t^2\right)\nonumber\\
&&\nonumber\\
  &&\!\!\!\!\!\!\!\!\!\!\!\!+\left(m_1^2-m_0^2-m_t^2\right)^2\Bigg\}B(m_t^2,
  m_0^2,m_1^2)-\left(m_1^2-m_t^2\right)^2B(m_t^2,0,m_1^2)\Bigg]\, .\nonumber
\end{eqnarray}

\end{appendix}

\end{document}